\documentclass[aps,pra,reprint,amssymb,superscriptaddress]{revtex4-2} 
\usepackage{amsmath,amssymb,amsfonts}
\usepackage[english]{babel}
\usepackage{graphicx}
\usepackage{dcolumn}
\usepackage{bm}
\usepackage[mathlines]{lineno}
\usepackage[colorlinks=true,citecolor=blue,linkcolor=red,urlcolor=blue]{hyperref}
\usepackage{orcidlink}

\usepackage{float}
\makeatletter
\let\newfloat\newfloat@ltx
\makeatother
\usepackage{algorithm}
\usepackage{algpseudocode}
\usepackage{dsfont}

\usepackage{braket}
\usepackage[dvipsnames]{xcolor}
\usepackage{booktabs}
\definecolor{tms}{rgb}{1,0,0}

\newcounter{sisection}

\usepackage{ragged2e}
\newcommand{\SRefSec}[1]{SI-\ref{#1}}

\newcounter{sifigure}

\newcommand{\SIFig}[1]{Fig.~\textcolor{red}{S}\hyperref[#1]{\ref{#1}}}
\newcommand{\SIFigs}[2]{Figs.~\textcolor{red}{S}\hyperref[#1]{\ref{#1}}--\textcolor{red}{S}\hyperref[#2]{\ref{#2}}}

\newcounter{edfigure}

\newcounter{edtable}

\newcommand{\newsection}[1]{%
\noindent\textbf{\fontsize{10.4}{12.4}\selectfont #1}\\
}
\usepackage{lineno}

\begin{document}

\title{One-to-one quantum simulation of a frustrated magnet with 256 qubits}


\author{Lucas Leclerc\textsuperscript{*}~\orcidlink{0000-0003-0581-9165}}
\affiliation{PASQAL SAS, 24 rue Emile Baudot, 91120 Palaiseau, France}

\author{Sergi Juli\`{a}-Farr\'e\textsuperscript{*}~\orcidlink{0000-0003-4034-5786}}
\affiliation{PASQAL SAS, 24 rue Emile Baudot, 91120 Palaiseau, France}

\author{Gabriel Silva Freitas\textsuperscript{*}~\orcidlink{0000-0001-6889-4897}}
\affiliation{Los Alamos National Laboratory (LANL), Los Alamos, New Mexico 87545, USA}

\author{Guillaume Villaret\textsuperscript{*}~\orcidlink{0000-0002-3898-8646}}
\affiliation{PASQAL SAS, 24 rue Emile Baudot, 91120 Palaiseau, France}

\author{Boris Albrecht~\orcidlink{0000-0003-0733-2676}}
\author{Lucas B\'{e}guin~\orcidlink{0000-0003-1388-0791}}
\author{Lilian Bourachot~\orcidlink{0009-0002-5860-9903}}
\author{Cl\'{e}mence Briosne-Frejaville~\orcidlink{0009-0007-2615-5050}}
\author{Dorian Claveau~\orcidlink{0009-0003-2694-1087}}
\author{Antoine Cornillot\,\orcidlink{0009-0004-7228-541X}}
\author{Julius de Hond\,\orcidlink{0000-0003-2217-934X}}
\author{Djibril Diallo~\orcidlink{0009-0007-8393-9329}}
\author{Cl\'{e}ment Dupays~\orcidlink{0009-0007-1944-7223}}
\author{Robin Dupont~\orcidlink{0009-0003-4541-5146}}
\author{Thomas Eritzpokhoff~\orcidlink{0009-0000-3899-8564}}
\author{Emmanuel Gottlob\,\orcidlink{0000-0003-3166-5497}}
\author{Lo\"ic Henriet~\orcidlink{0000-0003-3108-0595}}
\author{Michael Kaicher\,\orcidlink{0000-0001-7986-5127}}
\author{Lucas Lassabli\`ere~\orcidlink{0000-0001-8081-1054}}
\author{Arvid Lindberg~\orcidlink{0000-0001-8714-8662}}
\author{Yohann Machu~\orcidlink{0009-0007-6766-6439}}
\author{Hadriel Mamann~\orcidlink{0009-0002-3832-6471}}
\author{Thomas Pansiot~\orcidlink{0009-0001-2099-0043}}
\author{Julien Ripoll~\orcidlink{0009-0004-5282-9942}}
\affiliation{PASQAL SAS, 24 rue Emile Baudot, 91120 Palaiseau, France}
\author{Eun Sang Choi}
\affiliation{National High Magnetic Field Laboratory, Florida State University, Tallahassee, Florida 32310, USA}
\author{Adrien Signoles~\orcidlink{0000-0001-7822-9444}}
\author{Joseph Vovrosh~\orcidlink{0000-0002-1799-2830}}
\author{Bruno Ximenez~\orcidlink{0009-0006-8985-1355}}
\affiliation{PASQAL SAS, 24 rue Emile Baudot, 91120 Palaiseau, France}


\author{Vivien Zapf~\orcidlink{0000-0002-8375-4515}}
\affiliation{Los Alamos National Laboratory (LANL), Los Alamos, New Mexico 87545, USA}

\author{Shengzhi Zhang}
\affiliation{National High Magnetic Field Laboratory, Florida State University, Tallahassee, Florida 32310, USA}
\author{Haidong Zhou}
\affiliation{Department of Physics and Astronomy, University of Tennessee, Knoxville, Tennessee 37996-1200, USA}


\author{Minseong Lee~\orcidlink{0000-0002-2369-9913}}
\affiliation{Los Alamos National Laboratory (LANL), Los Alamos, New Mexico 87545, USA}
\author{Tiago Mendes-Santos~\orcidlink{0000-0001-6827-5260}}
\author{Constantin Dalyac~\orcidlink{0000-0002-0339-6421}}

\affiliation{PASQAL SAS, 24 rue Emile Baudot, 91120 Palaiseau, France}
\author{Antoine Browaeys~\orcidlink{0000-0001-9941-8869}}
\affiliation{PASQAL SAS, 24 rue Emile Baudot, 91120 Palaiseau, France}
\affiliation{Universit\'{e} Paris-Saclay, Institut d'Optique Graduate School, CNRS, Laboratoire Charles Fabry, 91127 Palaiseau Cedex, France}
\author{Alexandre Dauphin~\orcidlink{0000-0003-4996-2561}}

\affiliation{PASQAL SAS, 24 rue Emile Baudot, 91120 Palaiseau, France}

\maketitle

\textbf{Analog quantum simulators offer a powerful microscopic probe of quantum many-body systems, yet have largely been benchmarked against model Hamiltonians rather than real materials. Here, we use a 256-qubit Rydberg simulator to implement the effective Hamiltonian of the frustrated triangular-lattice magnet TmMgGaO$_4$. Simulated magnetization curves agree quantitatively with susceptibility measurements on single crystals, and both platforms consistently determine the antiferromagnetic phase transition. Snapshot-resolved analysis confirms that quantum fluctuations, rather than disorder, govern the intermediate paramagnetic regime. Having established this correspondence, we access non-equilibrium dynamics following a sudden quench, a regime at picosecond material timescales where entanglement growth places the problem beyond classical reach. The simulator reveals thermalization of local observables, demonstrating that analog quantum simulation can reproduce and extend the physics of a real material.}


Analogue quantum simulators~\cite{manin_computable_1980,feynman_simulating_1982,georgescu_quantum_2014,daley_twenty-five_2023}—naturally suited to large-scale, highly entangled systems—have so far been used primarily to demonstrate universal physical phenomena, including ground-state properties~\cite{scholl_quantum_2021,ebadi_quantum_2021,king_quantum_2021,narasimhan_simulating_2024,kairys_simulating_2020,mazurenko_cold-atom_2017,semeghini_probing_2021} and nonequilibrium dynamics~\cite{manovitz_quantum_2025,ali_quantum_2024}. Their potential for quantitative comparison with solid-state experiments and for predictive modelling of specific materials remains largely unexplored. Classical numerical techniques, by contrast, are routinely used to reproduce experimental observations and to explore new regimes in specific materials; however, their applicability is ultimately constrained by system size or entanglement. As analogue platforms reach sizes where classical numerics become increasingly challenging, while offering better control over noise and coherence, quantitatively connecting them to real materials is becoming both feasible and timely.

In this context, low-dimensional quantum materials are a natural test bed to validate quantum simulators and probe their predictive power. First explored theoretically in the mid-twentieth century~\cite{tomonaga_remarks_1950,anderson_approximate_1952,mermin_absence_1966}, these systems have since been realised across a wide range of platforms, including semiconductor heterostructures, bulk layered materials and synthetic two-dimensional systems~\cite{liu20192d,scappucci2021crystalline,pham20222d}. Understanding the microscopic mechanisms underlying their emergent properties has led to a a close interplay between solid-state experiments, analytical and classical numerical methods. Quantum simulators can now add a complementary role to this toolbox.

\begin{figure*}[t!]
    \centering
    \includegraphics[width=\textwidth]{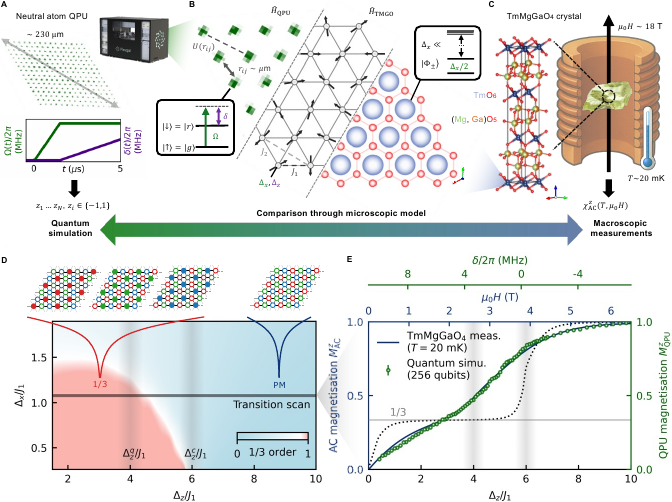}    \caption{\textbf{Bridging quantum simulation and macroscopic measurements of a frustrated magnet. }%
    \textbf{A,} Neutral-atom QPU with $N=256$ atoms, time-dependent adiabatic protocols $\Omega(t),\delta(t)$ and read out by site-resolved projective measurements $z_i$. %
    \textbf{B,} Mapping from a programmable quantum simulator to the microscopic description of TmMgGaO$_4$.
    (left) Effective Rydberg Hamiltonian with qubit states $\ket{g}$, $\ket{r}$ coupled by Rabi frequency $\Omega/2\pi$ and detuning $\delta/2\pi$, interacting via a distance-dependent potential $U(r_{ij})$.
    (centre) Microscopic model with easy-axis Ising interactions $J_1$ and $J_2$, and transverse and longitudinal fields $(\Delta_x,\Delta_z)$.
    (right) Two-dimensional triangular lattice of Tm$^{3+}$ ions; inset shows the effective pseudospin-$1/2$ doublet $\ket{\Phi_\pm}$ with crystalline-field induced splitting $\Delta_x/2$ and separated from higher energy levels.
    \textbf{C,} Sketch of experimental setup for thermodynamic measurements of AC susceptibility $\chi^z_{\rm AC}(\mu_0H,T)$ on a TmMgGaO$_4$ single-crystal sample in longitudinal magnetic fields up to $18\,\mathrm{T}$ and at temperatures down to $20\,\mathrm{mK}$. Zoom on the crystal internal structure. 
    \textbf{D,} Zero-temperature phase diagram $(\Delta_z/J_1,\Delta_x/J_1)$ from density-matrix-renormalisation group algorithm (see Supp. Mat.), showing paramagnetic (all spins up; hollow dots) and $1/3$-filling ordered (one sublattice with spins down; filled dots) phases; $\Delta_x/J_1$ of TmMgGaO$_4$~\cite{li_kosterlitz-thouless_2020} used for transition scan (solid black). Classical and quantum critical points estimates (gray). 
    \textbf{E,} (left axis) AC magnetisation $M^z_{\rm AC}(\Delta_z/J_1)$ at $T=20~\mathrm{mK}$ (blue) compared to QPU magnetisation $M^z_{\rm QPU}(\Delta_z/J_1)$ at $N=256$ (green dots); error bars reflect the finite number of QPU measurements. Magnetisation obtained from cut of DMRG phase diagram of $\textbf{D}$ in the quasi-classical regime $\Delta_x/J_1=0.2$ (dotted line). Horizontal axes map experimental ($\mu_0H$), microscopic-model ($\Delta_z/J_1$), and quantum simulator ($\delta/2\pi$) parameters.}
    \label{fig:fig1}
\end{figure*}

Here we investigate the weakly coupled magnetic planes of TmMgGaO$_4$~\cite{cevallos_anisotropic_2018}. We perform a direct comparison between macroscopic magnetic measurements in a solid-state experiment~\cite{dun_chemical_2014} and microscopic observations obtained using an analogue quantum simulator: a Rydberg-based quantum processing unit (QPU) manipulating up to 256 qubits~\cite{saffman_quantum_2010,browaeys_many-body_2020,henriet_quantum_2020}. TmMgGaO$_4$ is an attractive platform to benchmark quantum simulators for several reasons. First, its magnetic properties and spin Hamiltonian have been extensively characterized by multiple independent groups, with consistent results across experiments~\cite{li_partial_2020,shen_intertwined_2019,li_kosterlitz-thouless_2020,liu_intrinsic_2020,qin_field-tuned_2022}. This agreement provides strong confidence in both the crystal quality and the underlying microscopic model, establishing a reliable experimental benchmark. Therefore agreement between quantum simulation and experiment not only validates the material description but also serves as a stringent test of the quantum simulator itself. Second, the Hamiltonian is well suited for analogue quantum simulators as described in detail below. The spin Hamiltonian includes terms that generates quantum entanglement, enabling direct comparison between quantum and classical approaches. Finally, despite extensive characterization, the origin of its magnetic behavior - whether driven by disorder or intrinsic quantum effects - remains unresolved, making it an ideal problem for quantum simulation. 


Our approach, sketched in Fig.~\ref{fig:fig1}, realises the original vision of quantum simulators~\cite{feynman_simulating_1982}: the intersimulation of two distinct quantum systems whose characteristic length scales differ by several orders of magnitude. In particular, a lattice constant of $3\,\text{\AA}$ in the solid-state material~\cite{cevallos_anisotropic_2018} is rescaled to $10\,\mu$m in the quantum simulator~\cite{browaeys_many-body_2020}.

We first compare equilibrium magnetisation curves of the material measured at ultralow temperatures with those extracted from many-body ground states prepared in the quantum simulator, observing excellent agreement. We further independently estimate the quantum critical point, exploiting experimentally accessible macroscopic signatures in the material and microscopic many-body observables in the simulator. We then connect ground-state quantum fluctuations to integrated inelastic neutron scattering data reported in Ref.~\cite{qin_field-tuned_2022}. A thorough snapshot-based analysis of quantum correlations allows us to explain further rich features observed in the magnetic curves of the material, confirming the hypothesis of quenched disorder being substantially small in TmMgGaO$_4$~\cite{li_partial_2020,shen_intertwined_2019,li_kosterlitz-thouless_2020,liu_intrinsic_2020,qin_field-tuned_2022}. 

Having established this quantitative correspondence, we exploit the ultrafast coherent dynamics accessible in the Rydberg quantum simulator to predict the non-equilibrium response of TmMgGaO$_4$ following a sudden quench of its longitudinal magnetic field. This regime, characterised by rapid entanglement growth, is particularly challenging for classical numerical methods~\cite{vovrosh_simulating_2025,vovrosh_resource_2025}. Access to this dynamical window enables the exploration of more fundamental aspects of the system, including the effective thermalisation of a closed quantum many-body system~\cite{dalessio_quantum_2016}. Our results motivate future experimental investigations of TmMgGaO$_4$, which would require access to picosecond time scales. \\

\newsection{Magnetic properties through different probes}
We now describe our cross-platform experimental study. 
The microscopic Hamiltonian of TmMgGaO$_4$ can be described by a spin $1/2$ transverse field Ising model on a triangular lattice~\cite{li_partial_2020,liu_intrinsic_2020,li_kosterlitz-thouless_2020,shen_intertwined_2019} (Fig.~\ref{fig:fig1}\textbf{B}),
\begin{equation}
\frac{\hat{H}_{\text{TMGO}}}{\hbar}= J_1\!\sum_{\langle i,j \rangle}\hat \sigma^z_i\hat \sigma^z_j+J_2\!\!\sum_{\langle\langle i,j \rangle\rangle}\!\!\hat \sigma^z_i\hat \sigma^z_j +\sum_{i=1}^N \left(\Delta_x\hat\sigma^x_i- \Delta_z\hat \sigma^z_i\right),
\label{eq:ising_tmgo}
\end{equation}
where $J_1$($J_2\approx 0.05J_1$\cite{li_kosterlitz-thouless_2020}) is the nearest (next-nearest) neighbour interaction coupling, $\Delta_x\approx 1.08J_1$~\cite{li_kosterlitz-thouless_2020} the intrinsic transverse field strength, $\Delta_z$ the external longitudinal magnetic field strength, $\hat{\sigma}^\mu_i$ the Pauli matrices and $\hbar$ the reduced Planck constant.

We simulate this Hamiltonian with a Rydberg-based QPU up to $N=256$ qubits acting as spins (Fig.~\ref{fig:fig1}\textbf{A}, Fig.~\ref{fig:register_samples}, and SM). $^{87}\mathrm{Rb}$ atoms are trapped in a triangular array of optical tweezers and the qubits are encoded into the ground state $\ket{\uparrow}=\ket{g}$ and a highly excited Rydberg state $\ket{\downarrow}=\ket{r}$, coupled by an effective single-photon laser with time-dependent controls given by the Rabi frequency $\Omega(t)$ and detuning $\delta(t)$. Different pulse protocols allow us to probe both ground state and nonequilibrium physics. Crucially, interactions between Rydberg states have almost the same neighbour dependence as Eq.~\eqref{eq:ising_tmgo} (see SM). 

TmMgGaO$_4$ is a layered frustrated rare earth magnet in which Tm$^{3+}$ ions form a two-dimensional triangular lattice~\cite{li_kosterlitz-thouless_2020} with a large interlayer distance and a strong easy-axis anisotropy, as shown in Fig.~\ref{fig:fig1}\textbf{B-C}. A key observable is the average bulk magnetisation $M^z\equiv 1/N_\text{b}\sum_{i\in \textrm{bulk}}\langle \hat\sigma^z_i \rangle$ (see Supp. Mat.). 
Previous studies comparing numerics and experiment found an energy scale of $J_1/(2\pi)\approx 60~\textrm{GHz}$ ($\hbar J_1\approx 0.25\,\rm meV$) and $\Delta_z/J_1\approx 1.543\,\mu_0H~$(T)~\cite{li_kosterlitz-thouless_2020, li_partial_2020, shen_intertwined_2019}. 
We probe the macroscopic magnetic properties of a TmMgGaO$_4$ sample (Fig.~\ref{fig:crystal-picture}) through quasi static AC susceptibility ($\chi^z_\text{AC}$) measurements for different applied magnetic fields and temperatures (Fig.~\ref{fig:fig1}\textbf{C}, and SM). In particular, we focus on a wide range of parameters, with a scan up to $\mu_0H=18$ T and temperatures down to $T=20$ mK, allowing for the precise characterisation of the interplay between thermal and quantum fluctuations.

Embodying the original vision of analogue quantum simulation~\cite{feynman_simulating_1982}, the QPU faithfully reproduces the equilibrium response of TmMgGaO$_4$ across the range of longitudinal fields indicated in the phase diagram of Fig.~\ref{fig:fig1}\textbf{D}. The magnetisation $M^z_{\rm QPU}$ obtained from the QPU following a quasi-adiabatic state-preparation protocol quantitatively matches the magnetisation $M^z_{\rm AC}$ extracted from AC susceptibility measurements (see Supp. Mat.) at $T = 20\,\mathrm{mK}$ [$k_{\rm B}T/(\hbar J_1) = 7\times 10^{-3}$], as strikingly illustrated in Fig.~\ref{fig:fig1}\textbf{E}. This agreement demonstrates that the QPU, while realizing a highly rescaled Hamiltonian, $\hat{H}_{\text{QPU}}\approx \alpha_{\text{QPU}}\hat{H}_{\text{TMGO}}$ with $\alpha_\text{QPU}\approx 1.5\times 10^{-5}$, keeps its key Hamiltonian-specific properties intact.

The separation of physical scales, $\alpha_{\rm QPU}$, however, strongly influences the available degrees of control and the observables that can be accessed. On the one hand, TmMgGaO$_4$ can be characterised in the thermodynamic limit via susceptibility measurements with precise control over temperature and longitudinal field, but single-site (\AA) resolution and ultrafast dynamics ($1/J_1 \sim \mathrm{ps}$) remain challenging. On the other hand, the QPU operates with a large but finite number of sites, yet enables time- ($1/J_1 \sim \mu\mathrm{s}$) and site- ($\sim \mu$m) resolved measurements, allowing direct observation of coherent spin dynamics that are out of reach even for quantum annealing devices~\cite{king_quantum_2021}. We now combine these complementary probes to explore the low-energy quantum regime of $\hat{H}_{\rm TMGO}$.\\

\begin{figure*}[t]
    \centering
    \includegraphics[width=\textwidth]{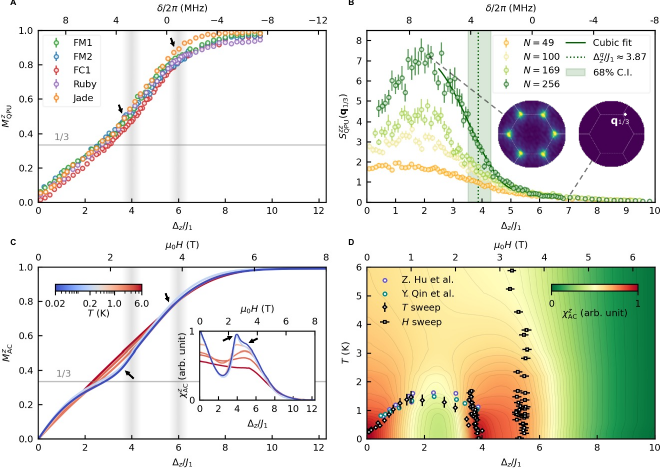}
    \caption{
    \textbf{Probing the paramagnet to $1/3$-order quantum phase transition.}
    \textbf{A,} QPU magnetisation $M^z_{\mathrm{QPU}}(\Delta_z/J_1)$ for $N=100$, measured on five devices : FM1 (green), FM2 (blue), FC1 (red), Ruby (purple) and Jade (orange dots); error bars reflect statistical noise from the finite number of measurements taken. \textbf{B,} Structure factor $S_{\rm QPU}^z(\mathbf{q}_{1/3})$ from FM1 data at $N=49,100,169,256$ (dots) with similar error bars; a cubic fit (solid line) for $N=256$ smooths the noise fluctuations to help estimating the location of the quantum phase transition (dashed) with $68\%$ confidence interval (shaded) obtained from varying fit windows. Insets: momentum-space maps of $S^{zz}_{\rm QPU}(\mathbf{q})$ across the transition for $|\mathbf{q}|\leq2\pi$.
    \textbf{C,} AC magnetisation $M^z_{\mathrm{AC}}(\Delta_z/J_1)$ of TmMgGaO$_4$ at decreasing $T$; inset: AC susceptibility $\chi^z_{\rm AC}$ highlighting emerging features (arrows). \textbf{D,} AC susceptibility $\chi^z_{\mathrm{AC}}(\Delta_z/J_1,T)$. Transition points extracted from field (squares) and temperature (diamonds) sweeps; previous results from \cite{hu_evidence_2020,qin_field-tuned_2022} shown for comparison. }
    \label{fig:fig2}
\end{figure*}

\newsection{Characterizing the quantum phase transition}
A finite size zero-temperature phase diagram of the model~\eqref{eq:ising_tmgo} is shown in Fig.~\ref{fig:fig1}\textbf{D}~\cite{liu_intrinsic_2020,da_liao_phase_2021,guo_order_2023}. In the absence of a transverse field, the system is classical and the transition occurs as a crossover ($\Delta_z^{\text{c}}/J_1\approx 6$) between the classical paramagnet with energy $E_{\uparrow\dots \uparrow}/N= 3J_1-\Delta_z+\mathcal{O}(J_2)$ and $M^z=1$, and the classical antiferromagnet with energy $E_{1/3}/N= -J_1-\Delta_z/3+\mathcal{O}(J_2)$ and $M^z=1/3$. A small finite transverse field leads to a quantum phase transition connecting the two classical $M^z$ plateaus (dotted line in Fig.~\ref{fig:fig1}\textbf{E}). The behaviour of $M^z_\textrm{AC}$ and $M^z_\text{QPU}$ can be interpreted as a strong quantum limit: a $1/3$ magnetisation elbow is still present, but the non-perturbative effect of quantum fluctuations ($\Delta_x=1.08J_1$) strongly shifts the critical longitudinal field and prevents the appearance of quasi-classical $M^z$ plateaus.

We probe the quantum phase transition using microscopic quantum simulation of $\hat{H}_{\rm TMGO}$ with the QPU. Quasi-adiabatically prepared ground states (SM, Fig.~\ref{fig:pulse_samples}) on $N=100$ sites across five independent, nearly identical QPUs (FM1, FM2, Ruby in France, Jade in Germany and FC1 in Canada) exhibit a subtle $1/3$ magnetisation elbow around $\Delta_z^{\rm q}/J_1 \approx 4$ and a broader feature near $\Delta_z^c/J_1 \approx 6$ (Fig.~\ref{fig:fig2}\textbf{A}). However, the minor effect of the critical point in the magnetisation curve makes locating it particularly challenging (SM, Fig.~\ref{fig:noise_study}). We turn to another observable accessible on the quantum simulator and examine the behaviour of the structure factor $S^{zz}_{\rm QPU}(\mathbf{q}) = 1/N_\text{b}\sum_{i,j\in \textrm{bulk}} e^{i \mathbf{q} \cdot \mathbf{r}_{ij}} \left[\langle \hat{\sigma}_i^z \hat{\sigma}_j^z \rangle - \langle \hat{\sigma}_i^z \rangle \langle \hat{\sigma}_j^z \rangle\right]$ in lattices up to $N=256$ sites on FM1 (Fig.~\ref{fig:fig2}\textbf{B}). A first feature is observed for $\Delta_z^c/J_1 \lesssim 6$, where a finite $S^{zz}_{\rm QPU}(\mathbf{q}_{1/3})$ signals the presence of quantum correlations. Long-range $1/3$ order for $\Delta^z/J_1<4$ is signalled by the emergence of Bragg peaks at $\mathbf{q}_{1/3}=2\pi/3\,(1,\sqrt{3})$ (see Fig.~\ref{fig:obs-scans} for a real-space visualisation). For moderately small systems ($N=49,\,100$), $S^{zz}_{\rm QPU}(\mathbf{q}_{1/3})$ already develops a broad peak within the $1/3$ phase, signalling the onset of this magnetic order. Larger-scale simulations ($N=169,\,256$) reproduce this feature with improved convergence in system size. The QPU-derived critical point estimate $\Delta_z^q/J_1(N=256) = 3.87_{-0.36}^{+0.44}$ (see SM). 

We next probe the same quantum phase transition on the single crystal.
In Fig.~\ref{fig:fig2}\textbf{C} we show how this magnetisation elbow emerges in the $M^z_\text{AC}$ curves obtained from $\chi^z_\text{AC}$ measurements (inset) at temperatures from $6$ K to $20$mK, in close agreement with previous measurements reaching $40\,$mK~\cite{li_partial_2020,shen_intertwined_2019,qin_field-tuned_2022}. We use the associated $\chi^z_\text{AC}$ peak to build the phase diagram of the system (Fig.~\ref{fig:fig2}\textbf{D}). We identify the quantum phase transition from the maximum of $|\partial^2 \chi^z_\text{AC} / \partial \Delta_z^2|$, at $\Delta_z^{\rm q}(20\,\mathrm{mK})/J_1 = 3.88 \pm 0.12$ (Fig.~\ref{fig:chi_ac}). The susceptibility peak grows systematically upon cooling, as expected. This scan also reveals a second intriguing feature near the classical critical point, $\Delta_z^c/J_1 \approx 6$, where a broad thermal peak at high temperatures gradually evolves into a low-temperature hump, reflecting the interplay of quantum and thermal fluctuations. Finally, we also perform temperature scans at fixed longitudinal fields (Fig.~\ref{fig:chi_ac}) to delimit the thermal dome of the $1/3$ phase in Fig.~\ref{fig:fig2}\textbf{D}. 

The QPU-derived critical point estimate is in good agreement with the value obtained from TmMgGaO$_4$ measurements, confirming the consistency of both approaches. In particular, the increase of $S^{zz}_\text{QPU}(\mathbf{q}_{1/3})$ with $N$ highlights the low scaling of noise effects with system size in analogue QPUs~\cite{trivedi_quantum_2024} and the robustness of quasi-adiabatic protocols. These are ultimately limited to study ground-state transitions in finite-time protocols by diabatic effects mostly arising from the many-body gap closing in the large $N$ limit. Numerical matrix product-state (MPS) emulations at $N=100$ benchmark the impact of finite preparation time with and without noise, supporting the QPU observations (SM, Fig.~\ref{fig:noise_study}). Furthermore, independent low-temperature Quantum Monte Carlo (QMC) simulations of $\hat{H}_{\rm QPU}$ (SM, Fig.~\ref{fig:QMC_data}) yield a comparable transition point and demonstrate that system sizes $N\gtrsim100$ are required to clearly resolve the quantum critical point.\\

\newsection{Quantum correlations and their connection with inelastic neutron scattering}
The low-temperature feature emerging at $\Delta_z^c/J_1 \approx 6$, observed both in QPU measurements (Fig.~\ref{fig:fig2}\textbf{A,B}) and in TmMgGaO$_4$(Fig.~\ref{fig:fig2}\textbf{C,D}), suggests an intermediate crossover regime where quantum correlations may play a significant role. While quenched-disorder–driven paramagnetic phases were initially proposed to explain this behaviour~\cite{li_partial_2020}, subsequent inelastic neutron scattering experiments appear inconsistent with that picture~\cite{shen_intertwined_2019,li_kosterlitz-thouless_2020,qin_field-tuned_2022}.

\begin{figure}[ht]
    \centering
    \includegraphics[width=0.5\textwidth]{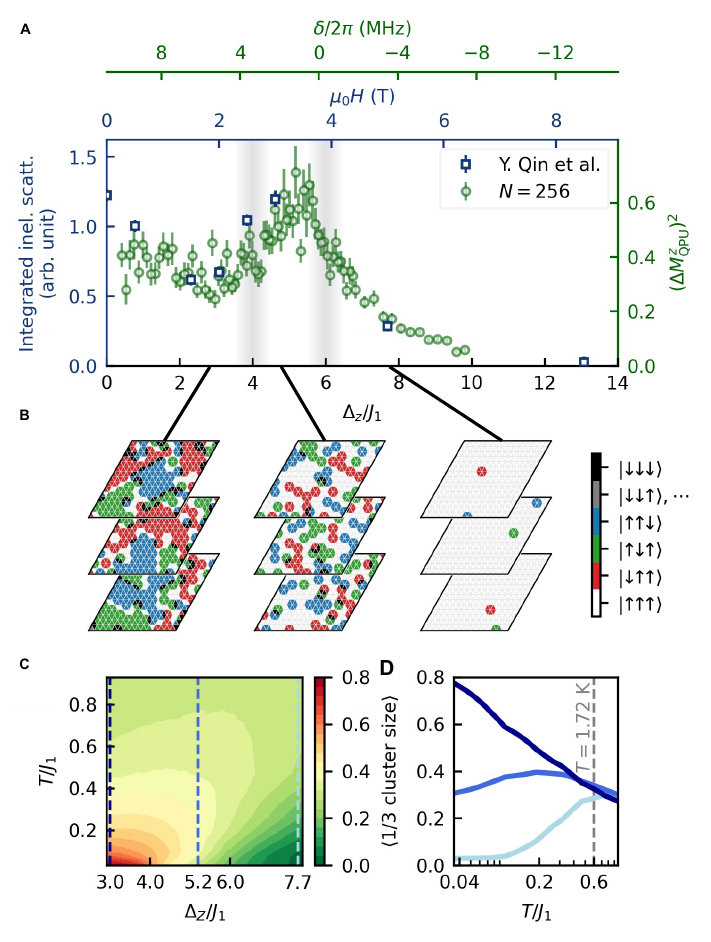}
    \caption{\textbf{Quantum fluctuations and emergence of $1/3$ order.}
\textbf{A,} (left axis) Low-temperature integrated inelastic neutron scattering signal from \cite{qin_field-tuned_2022} (blue squares), given by the difference between total scattering at $T=0.13\,\mathrm{K}$ and elastic scattering at $T=40\,\mathrm{mK}$. (right axis) Variance of the QPU magnetisation $(\Delta M^z_{\mathrm{QPU}})^2 $ as a function of applied field $\Delta_z/J_1$ for $N=256$. QPU measurements (green dots) with error bars reflecting statistical noise from the finite number of shots.
\textbf{B,} Representative single-shot spin configurations from the QPU, with plaquettes coloured by triangle configuration. \textbf{C,} Phase diagram of the mean $1/3$-cluster size versus temperature $T/J_1$ (we set $k_\text{B}/\hbar=1$) and field $\Delta_z/J_1$ from QMC at $N=256$. \textbf{D,} Vertical cuts from \textbf{C} at $\Delta_z/J_1=3,~5.2,~7.7$ (light to dark blue).}
    \label{fig:fig3}
\end{figure}

To probe these correlations, we exploit a connection~\cite{scheie_tutorial_2025} between the integrated inelastic neutron scattering signal and fluctuations of the global magnetisation, $(\Delta M^z_{\rm QPU})^2 = \langle (\hat{M}^z_{\rm QPU} - \langle \hat{M}^z_{\rm QPU} \rangle)^2 \rangle$, which is directly accessible in the QPU. Comparing $(\Delta M^z_{\rm QPU})^2$ with the integrated signal from Ref.~\cite{qin_field-tuned_2022} (Fig.~\ref{fig:fig3}\textbf{A}) reveals qualitative agreement, highlighting the simulator’s ability to capture correlations present in the material.

Moreover, for a pure quantum state, $(\Delta M^z_{\rm QPU})^2$ is proportional to the quantum Fisher information, which provides a lower bound on multipartite entanglement~\cite{toth_multipartite_2012,hyllus_fisher_2012}. The observed enhancement of this quantity at low temperatures in both the material and the QPU indicates that the intermediate paramagnetic regime preceding the $1/3$ quantum phase transition is also primarily governed by quantum fluctuations rather than strong quenched disorder.\\

\newsection{Microscopic nature of the quantum paramagnet}
At $\Delta_z/J_1 \approx 6$, the classical crossing between the energies of the classical configurations, namely $E_{\uparrow\dots \uparrow}$ and $E_{1/3}$, also leads to an extensive degeneracy, arising from gapless local $1/3$ excitations in $\ket{\uparrow \dots \uparrow}$. In this picture, also discussed in Ref.~\cite{king_quantum_2021}, a finite transverse field $\Delta_x$  generates a paramagnetic superposition of these excitations, preserving the gap finite until the quantum phase transition occurs at a lower $\Delta_z/J_1$. 

This superposition is consistent with the enhanced magnetisation fluctuations, $(\Delta M^z_{\rm QPU})^2$, observed in Fig.~\ref{fig:fig3}\textbf{A}, and microscopic snapshots from the QPU further illustrate this behaviour (Fig.~\ref{fig:fig3}\textbf{B}). Within the peak region, bitstrings exhibit local $1/3$-ordered patches superimposed on the fully polarised background, whereas outside this region the system displays large-domain $1/3$ order or nearly fully polarised $z$-alignment. Interestingly, the regime where this paramagnetic regime with enhanced quantum fluctuations and short-range $1/3$ order appears corresponds, in TmMgGaO$_4$, to the region between the $\chi^z_\text{AC}$ peak and hump (inset of Fig.~\ref{fig:fig2}\textbf{D}, Fig.~\ref{fig:chi_ac}). The QPU microscopic analysis thus strongly suggests that the peak-hump duality in the susceptibility curve (inset of Fig.~\ref{fig:fig2}\textbf{C}) is a hallmark of a dominant transverse field term $\Delta_x$ without requiring auxiliary mechanisms, e.g., quenched disorder. 


Finally, from the TmMgGaO$_4$ susceptibility curves measured at different temperatures, we identify a regime $T \gtrsim 1$~K in which the peak-hump structure evolves smoothly into a single broad thermal peak located at $\Delta_z/J_1 \approx 6$. We investigate this regime using equilibrium QMC sampling (see Supp. Mat.), which allows us to compute the average size of clusters exhibiting $1/3$ order (Fig.~\ref{fig:fig3}\textbf{C,D}). Although the clusters remain short-ranged overall, their size displays a pronounced non-monotonic temperature dependence: short-range $1/3$ order initially grow upon heating before being suppressed again at higher temperatures. This behaviour, absent at both lower and higher fields, further supports a dominant role of the quantum term $\Delta_x$~\cite{king_quantum_2021}: thermal activation across the paramagnetic gap, of order $\hbar\Delta_x/k_\mathrm{B} \approx 3.13$~K, enables the population of low-lying excited states with strong $1/3$ character, enhancing correlations at intermediate temperatures due to a quantum-thermal interplay. \\

\newsection{Post-quench dynamics and thermalisation}
We now turn to the non-equilibrium dynamics following an abrupt change, or quench, of the longitudinal magnetic field that can give insights into the energy spectrum through spectroscopy, quasiparticle excitations and thermalisation. Such post-quench dynamics are, in principle, experimentally accessible in real materials: time-resolved magneto-optical measurements could directly monitor the evolution of the magnetisation, while ultrafast pump–probe spectroscopies and neutron scattering could probe relaxation. However, the relevant microscopic timescales in these materials are typically on the order of picoseconds, placing these experiments beyond the scope of the present work.

From a computational perspective, accurately simulating post-quench dynamics in two dimensions is exceptionally challenging for classical methods~\cite{vovrosh_simulating_2025,vovrosh_resource_2025}. In contrast, these dynamics are naturally accessible on a QPU~\cite{hyosub_detailed_2018}. The corresponding physical timescales are of order microseconds, with achievable time resolution well below $0.1\mu\mathrm{s}$. Importantly, exploring this dynamical regime requires no modifications to the experimental platform beyond those already employed for the equilibrium and near-adiabatic protocols described above, making post-quench dynamics an especially compelling target of quantum simulation~\cite{quenchpaper}.

\begin{figure}[ht]
    \centering
    \includegraphics[width=\linewidth]{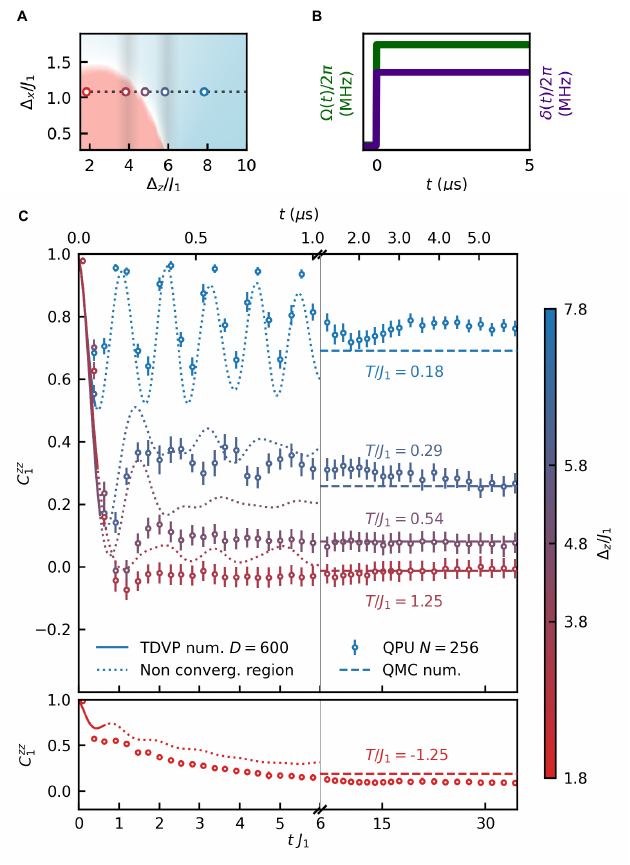}
    \caption{\textbf{Thermalisation of post-quench dynamics.} \textbf{A,} Phase diagram (cf. Fig.~\ref{fig:fig1}\textbf{D}) indicating quenches performed across the phase transition. \textbf{B,} Time-dependent quench protocols programmable on the QPU. \textbf{C,} Post-quench dynamics of the nearest-neighbour correlation $C_1^{zz}$ for different $\Delta_z/J_1$ at $N=256$ measured on the QPU (dots; error bars reflect statistical noise from the finite number of shots). Short-time dynamics ($tJ_1=\mathcal{O}(1)$) are compared to TDVP-MPS simulations ($D=600$; solid: converged, dotted: unconverged), while long-time dynamics ($tJ_1=\mathcal{O}(10)$) are compared to thermalised QMC results (dashed) at $T/J_1$ (we set $k_\text{B}/\hbar=1$).}
    \label{fig:fig.4}
\end{figure}

We now demonstrate the flexibility of the QPU by probing post-quench dynamics of a large-scale $N=256$ system, as shown in Fig.~\ref{fig:fig.4}\textbf{A}. Starting from the paramagnetic product state $\ket{\uparrow \dots \uparrow}$, we abruptly change the Hamiltonian parameters (Fig.~\ref{fig:fig.4}\textbf{B}) and monitor the ensuing non-equilibrium evolution.

We first benchmark the short-time dynamics against classical simulations based on MPS evolved using the time-dependent variational principle (TDVP) with bond dimension $D=600$ (Fig.~\ref{fig:fig.4}\textbf{C}). Specifically, we compute the nearest neighbour correlator  $C^{zz}_1\equiv 1/N_b^{\prime} \sum_{\langle i,j\rangle\in \textrm{bulk}}\langle\hat{\sigma}^z_i\hat{\sigma}^z_j\rangle$ (where $N_b^{\prime}$ is the number of bonds $\langle i,j\rangle\in \textrm{bulk}$) and observe good quantitative agreement at early times ($tJ_1\leq1$), where the MPS results are converged (\SRefSec{si-sec:resource estimation}). For large $\Delta_z/J_1$, the oscillation frequency observed in both approaches can be directly related to the gap of the paramagnetic phase. In contrast, for smaller values of $\Delta_z/J_1$, the dynamics become increasingly damped and no single characteristic frequency can be identified, consistent with the breakdown of a simple quasiparticle picture.
At longer times, however, the dynamics generate increasing amounts of entanglement, requiring progressively larger bond dimensions to maintain convergence. Since the computational cost of TDVP scales steeply with bond dimension and system size—specifically, with runtime $\sim O(N^2 D^3)$ and memory $\sim O(N^{3/2}D^2)$—simulations of large systems quickly become prohibitive. For the parameters considered here, the corresponding MPS simulations require roughly two weeks of wall-clock time, compared to about one day on the QPU. A detailed analysis of resource requirements is provided in the Supplementary Materials and Fig.~\ref{fig:resource_estimates}.

When interpreting these experimental results, several points should be kept in mind. First, in the quenching experiments the spin system is assumed to evolve in an effectively adiabatic regime, largely decoupled from the phonon bath on experimental timescales. This assumption is consistent with the low-temperature regime, where phonons with energies comparable to the spin excitations are strongly suppressed, leaving primarily low-energy acoustic modes that are inefficient for energy exchange. As a result, the spin–phonon relaxation time becomes significantly prolonged. Second, inelastic neutron scattering measurements at low temperatures do not show clear evidence of strong spin–phonon coupling channels~\cite{qin_field-tuned_2022}, supporting that such coupling is strongly suppressed in TmMgGaO$_{4}$ in this regime.

To characterise the long-time regime, we probe whether local observables of the evolving state thermalise~\cite{dalessio_quantum_2016}. In particular, expectation values measured on the QPU are compared with those of an equilibrium canonical ensemble for $\hat{H}_{\textrm{QPU}}$ at an effective temperature $T/J_1$, which is defined under the assumption of an isolated system  during the energy-conserving unitary dynamics.
The energy of the initial state $\ket{\uparrow \dots \uparrow}$ is matched to that of the thermal ensemble using QMC simulations~\cite{sandvik_sthocastic_2003,merali_stochastic_2024} (see Supp. Mat.). Remarkably, correlations measured on the QPU agree closely with the thermal predictions (Fig.~\ref{fig:fig.4}\textbf{C}), providing evidence of thermalisation. This correspondence validates QPU results at system sizes and evolution times beyond the reach of classical TDVP simulations, and highlights the open question of whether alternative classical methods can reproduce these long-time dynamics.\\

\section*{Conclusions and outlook}
In this work, we demonstrated that programmable two-dimensional arrays of neutral atoms with Rydberg-state interactions provide a faithful quantum simulation of the two-dimensional quantum magnet TmMgGaO$_4$. We observed quantitative agreement between macroscopic measurements of the material and results obtained on the Rydberg platform across its equilibrium quantum phase transition. The simulator also enabled the study of strongly correlated spin dynamics out of equilibrium, highlighting its operation in a regime of site-resolved measurements and high temporal resolution—regimes that are challenging to access in real materials or with classical simulations.

While we focused here on the antiferromagnetic quantum phase transition, our approach—combining macroscopic sample measurements with quantum simulation—can be extended to explore other characteristic phenomena of TmMgGaO$_4$, such as the exotic clock phase, with six-fold degeneracy, at $\Delta_z/J_1 = 0$~\cite{li_kosterlitz-thouless_2020,fey_quantum_2019,moessner_ising_2001}.

More broadly, this work illustrates that current neutral-atom devices can be employed to simulate magnetic materials following a systematic quantum simulation paradigm: an initial model-certification stage compares macroscopic observables with the QPU, a microscopic analysis probes the underlying physics, and a final prediction stage explores physical regimes that are largely inaccessible to conventional solid-state experiments and classical numerical methods.  

This strategy opens the door to studying both equilibrium and nonequilibrium phenomena across a wide class of quantum magnets, particularly layered or synthetic two-dimensional systems with easy-axis Ising interactions. Recent advances in Rydberg platforms—enabling three-dimensional arrays and the implementation of more general spin Hamiltonians, including XY~\cite{chen_continuous_2023} and XYZ~\cite{scholl_microwave_2022} interactions—further expand the range of quantum materials that can be faithfully simulated and explored.
\bibliography{bibliography/tmgo_bib,bibliography/maglab}

\section*{Data and materials availability}
The data supporting the findings of this study are available on reasonable request.

\section*{Acknowledgments}
We thank Carleton Coffrin and Christophe Jurczak for the preliminary discussions that led to this work. We thank Mourad Beji and Louis-Paul Henry for carefully reading the manuscript and providing insightful suggestions. More generally, we thank the various teams at Pasqal making this project possible, including the Hardware and Software platform departments, the Emulator team and Pasqal Canada Inc.
\section*{Funding}
Pasqal acknowledges funding from the European Union under the projects PASQuanS2.1 (HORIZON-CL4-2022-QUANTUM02-SGA, Grant Agreement 101113690). Pasqal acknowledges the usage of the Ruby and Jade machines, the installation of which was supported by the European High-Performance Computing Joint Undertaking (JU) under grant agreement No 101018180 and project name HPCQS. AB also acknowledges support by the Agence Nationale de la Recherche (ANR-22-PETQ-0004 France 2030, project QuBitAF). GSF, VZ, and ML acknowledge the LANL LDRD program for funding the scientific work. Magnetic measurements were performed at the National High Magnetic Field Laboratory, which is supported by National Science Foundation Cooperative Agreement No.\ DMR-2128556, the State of Florida, and the U.S.\ Department of Energy. The work at the University of Tennessee (crystal growth) is supported by the National Science Foundation under Grant No.\ DMR-2003117 to HZ. Pasqal team acknowledge AWS for their support and resources for running large-scale numerical simulations.
\section*{Author contributions}
LL, SJF, GSF, and GV contributed equally to this work. GV, DC, and LB designed and executed the Rydberg quantum simulation experiments. GSF, ML, SZ, and ESC performed the material characterisation and magnetisation measurements at the magnetic laboratory facilities. HZ synthesised and provided the TmMgGaO$_4$ single crystals. LL, SJF, JV, TMS, EG, and MK developed the theoretical models, mapping protocols, and conducted numerical simulations. YM and CBF engineered the expanded optical field-of-view for the atom arrays. JR, JH, BA, and TE developed and implemented the QPU calibration and characterisation sequences. AC, HM, BX, and RD developed the protocols for large-scale, low-defect atom array assembly. TP and DD designed and implemented the critical control software and operating system architecture. CoD, AD, LH, ML, AS, VZ, and AB provided scientific leadership and supervised the project. All authors participated in the analysis of the results and contributed to the final manuscript.
\section*{Competing interests}
LL, SJF, GV, DC, AC, LB, BA, CBF, YM, HM, JR, JH, BX, ClD, TP, DD, RL, AL, RD, LL, JV, TMS, EG, MK, CoD, LH, AS, AB and AD are employees and/or shareholders of Pasqal. The remaining authors declare no competing interests.

\clearpage
\onecolumngrid
\renewcommand{\thesection}{S\arabic{section}}
\renewcommand{\thefigure}{S\arabic{figure}}
\renewcommand{\thetable}{S\arabic{table}}
\renewcommand{\theequation}{S\arabic{equation}}
\setcounter{equation}{0}
\setcounter{figure}{0}
\setcounter{table}{0}
\setcounter{section}{0}

\begin{center}
    {\Large\bfseries Supplementary Materials for}\\[0.5em]
    {\Large\bfseries One-to-one quantum simulation of a frustrated magnet with 256 qubits}\\[1.5em]
\end{center}

\vspace{2em}
\hrule
\vspace{1em}
\section*{Contents}
\begin{enumerate}
    \item \textbf{Supplementary Note 1:} Hardware and Platform Integration ..................................................... Page \pageref{note:hardware}
    \item \textbf{Supplementary Note 2:} Experimental Procedures ..................................................................... Page \pageref{note:procedures}
    \item \textbf{Supplementary Note 3:} Theoretical Mapping ............................................................................ Page \pageref{note:mapping}
    \item \textbf{Supplementary Note 4:} Material Synthesis and Characterization ............................................. Page \pageref{note:material}
    \item \textbf{Supplementary Note 5:} Numerical Methods, Noise Model and Resource Estimations .............. Page \pageref{note:numerics}
\end{enumerate}
\vspace{1em}
\hrule
\vspace{2em}

\section*{Supplementary Note 1: Hardware description and platform integration} \label{note:hardware}

The quantum simulations were performed on five Orion Beta devices (FM1, FM2, FC1, Jade and Ruby), the latest Pasqal quantum processing units based on programmable arrays of Rydberg atoms~\cite{saffman_quantum_2010,browaeys_many-body_2020,henriet_quantum_2020}. Unless otherwise specified, all references to QPUs in the text refer to the FM1 device. Located in France, FM1 was used to acquire the majority of the data presented in the main text and operates with atoms excited to the Rydberg state $\ket{r}=\ket{75S_{1/2},m_J=1/2}$, supporting arrays of up to $256$ atoms. Four additional Orion Beta devices (100qubits) were employed for cross-checks shown in Fig.~\ref{fig:fig2}\textbf{B}: FM2 (France), FC1 (Canada), Ruby (France) and Jade (Germany). The integration of Jade and Ruby QPUs into the HPC infrastructures of FZJ and CEA within the HPCQS project showcases the reliability of the neutral-atom platform, proving its capacity for autonomous operation in a high-availability supercomputing environment. Although these QPUs operate at a different Rydberg level, we rescale the lattice spacing of the atomic arrays to realise the same effective microscopic Hamiltonian across devices. Aside from these adjustments, the Orion Beta platforms share very similar technical characteristics; further device-specific details are summarised in the Supplementary Table~\ref{tab:qpu_comparison}.\\


\renewcommand{\tablename}{Supplementary Table}

\begin{table*}[h]
\small
\centering
\begin{tabular*}{\textwidth}{@{\extracolsep{\fill}} l ccccc @{}}
\toprule
 & \textbf{FM1} & \textbf{FM2} & \textbf{FC1} & \textbf{Ruby} & \textbf{Jade} \\
\midrule

Location 
& Massy (FR) 
& Massy (FR) 
& Sherbrooke (CA) 
& CEA, TGCC (FR) 
&  Jülich FZJ (DE) \\

$N_{\text{max}}$ 
& 256 & 100 & 100 & 100 & 100 \\

Rydberg level 
& $|75S_{1/2},m_J=1/2\rangle$ 
& $|60S_{1/2},m_J=1/2\rangle$ 
& $|60S_{1/2},m_J=1/2\rangle$ 
& $|60S_{1/2},m_J=1/2\rangle$  
& $|60S_{1/2},m_J=1/2\rangle$  \\

Spacing $r_1$ ($\mu$m) 
& 9 & 5.8 & 5.8 & 5.8 & 5.8 \\

False pos. $\varepsilon$ (\%) 
& 1 & 1 & 1 & 1 & 1 \\

False neg. $\varepsilon'$ (\%) 
& 4 & 7 & 7 & 7 & 7 \\

Adiabatic ramp 
& AOM/EOM 
& AOM/AOM 
& AOM/AOM 
& AOM/AOM 
& AOM/AOM \\

Traps off 
& Variable (post-EOM) 
& Fixed (6 $\mu$s) 
& Fixed (6 $\mu$s) 
& Fixed (6 $\mu$s) 
& Fixed (6 $\mu$s) \\

Post-quench ramp 
& EOM/EOM 
& --- 
& --- 
& --- 
& --- \\

\bottomrule
\end{tabular*}
\caption{Comparison of the specifications and usage of the various QPUs used in this work.}
\label{tab:qpu_comparison}
\end{table*}

\clearpage
\section*{Supplementary Note 2: experimental procedures} \label{note:procedures}
\textbf{Preparation of triangular atomic arrays}

During the initialisation of the atomic array, up to approximately $275$ individual $^{87}\mathrm{Rb}$ atoms are laser-cooled and stochastically loaded into optical tweezers at 815~nm generated by a spatial light modulator (SLM). The atoms are subsequently rearranged into deterministic patterns with tunable spacings of around $10~\mu$m using a single moving optical tweezer operating at 850~nm. The probability of preparing defect-free arrays decreases rapidly with system size. To further increase the repetition rate of our computations up to $256$ qubits, we implement two rearrangement steps and apply post-selection of defects. While some applications may require strictly defect-free arrays, in this work an optimum is sought by balancing the fraction of tolerable defects against the achievable repetition rate. This ensures both efficient data acquisition and manageable defect-induced effects. More specifically, we discard shots in which the average defect number exceeds $2\%$. This procedure yields a typical repetition rate of $0.7\,\mathrm{Hz}$ for 256-qubit arrays. For remaining shots with defects, sites with missing atoms are explicitly identified in the experimental readout (X instead of $0/1$) to allow for further bitstrings post-processing.
\begin{figure}
    \centering
     \includegraphics[width=0.7\textwidth]{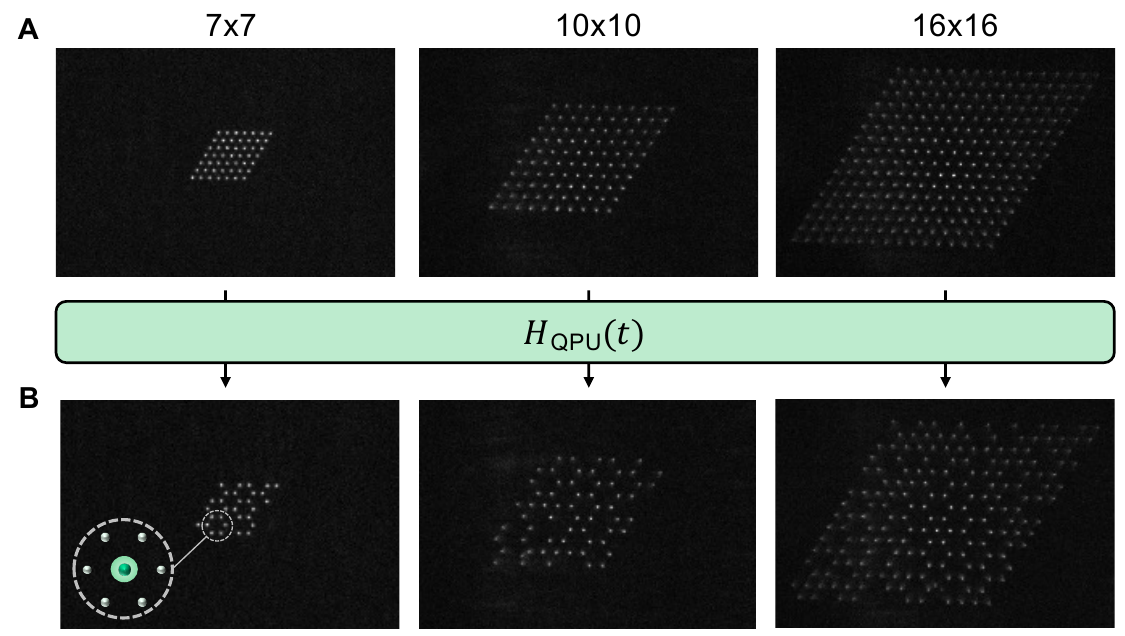}
     \caption{\textbf{Fluorescence images of triangular-lattice atomic registers used in the QPU simulations.} 
\textbf{A,} Images acquired after atom rearrangement, prior to the quantum simulation for $N=49,100$ and $256$. 
\textbf{B,} Images acquired at the end of the $1/3$-phase preparation. For clarity, bitstrings with large $1/3$-ordered domains are shown.
}
     \label{fig:register_samples}
\end{figure}

An additional laser-cooling stage is then applied followed by optical pumping under a 10~G magnetic field, preparing $8~\mu$K cooled atoms in $120~\mu$K deep traps. All atoms are thus prepared in the ground state $\ket{g}=\ket{5S_{1/2},\,F=2,\,m_F=2}$. After this preparation stage, the optical traps are turned off for the coherent evolution stage (i.e., the quantum simulation protocol) described in the next sections. 

To realize triangular arrays suitable for the preparation of the $1/3$ antiferromagnetic phase of the model, atoms are arranged in an $N=L \times L$ rhombus geometry, as shown in the fluorescence images of Fig.~\ref{fig:register_samples}. Commensurability of the $1/3$ phase requires that $L$ be a multiple of three in a periodic system, ensuring that the desired $1/3$-filling order shown in Fig.~\ref{fig:fig1}\textbf{D} can fit perfectly within the finite array. To mitigate edge effects inherent to systems with boundaries, the bulk of interest is surrounded by additional two rows of atoms, yielding total array lengths of $L=7, 10, 13$ or $16$ in the experiments. While different than previous works preparing the $1/3$ phase \cite{scholl_quantum_2021}, this choice of rhombus geometry balances the need for commensurability with the suppression of boundary-induced distortions.\\

\textbf{Rydberg Hamiltonian on atomic arrays}

Each local lattice qubit is encoded in the ground state $\ket{g}$ and a highly excited Rydberg state $\ket{r} = \ket{nS_{1/2}, m_J=1/2}$ with principal quantum number $n=75$ on FM1. The two states are coupled via a standard two-photon transition~\cite{browaeys_many-body_2020,scholl_quantum_2021-1} through the $6P_{3/2}$ intermediate state using counter-propagating lasers at 420~nm and 1013~nm, with intermediate-state detuning $\Delta/(2\pi) = 400~\mathrm{MHz}$ and Rabi frequencies $\Omega_{\rm 420}$ and $\Omega_{\rm 1013}$. Cylindrical lenses with an aspect ratio of 4 enhance the laser intensity at the atoms while using standard laser powers (1~W cavity-doubled diode at 420~nm, 10~W fiber-amplified diode at 1013~nm). Adiabatic elimination of the intermediate state~\cite{browaeys_many-body_2020,scholl_quantum_2021-1} results in an effective transition between $\ket{g}$ and $\ket{r}$, defined by its Rabi frequency $\Omega$ and detuning $\delta$. Acousto-optical modulators (AOM) allow to modulate the intensity of both lasers, as well as the frequency of the 1013~nm laser, and thus create arbitrary temporal waveforms for $\Omega(t)$ and $\delta(t)$ with typical rise times of $\sim 80 \, \text{--} \, 100 \, \mathrm{ns}$. An additional electro-optical modulator (EOM) on the 420~nm laser beamline allow to shape square pulses for $\Omega(t)$ with a faster rise time of $\sim 20 \; \text{ns}$. The resulting quantum simulator Hamiltonian reads~\cite{browaeys_many-body_2020, scholl_quantum_2021-1}
\begin{equation}
\label{eq:ham_qpu_1}
\hat{H}_{\rm QPU} = \sum_{i<j} U_{ij} \hat n_i \hat n_j + \frac{\hbar\Omega(t)}{2} \sum_i \hat \sigma_i^x - \hbar\delta(t) \sum_i \hat n_i,
\end{equation}
where $\hat n_i \equiv (1 - \hat \sigma_i^z)/2$ counts the local Rydberg population. In this representation, $\sigma_i^z\ket{g}=+ \ket{g}$ and $\sigma_i^z\ket{r}=- \ket{r}$. $U_{ij} = C_6(n)/r_{ij}^6$ is the van der Waals interaction between excitations. Typical values for nearest-neighbour distances are $R_{\rm 1}=9~\mu$m, leading to $U_{\rm 1}/(2\pi\hbar)=3.7$ MHz with $C_6(n=75)/(2\pi\hbar)\approx1949$ GHz$\cdot\mu$m$^6$. The available laser powers allow to reach Rabi frequencies $\Omega(t)/(2\pi)$ up to $2 \, \mathrm{MHz}$, while the detuning $\delta(t)/(2\pi)$ can be scanned over the $\left[ -14, +14 \right] \, \mathrm{MHz}$ range.\\

\begin{figure}
    \centering
     \includegraphics[width=0.5\linewidth]{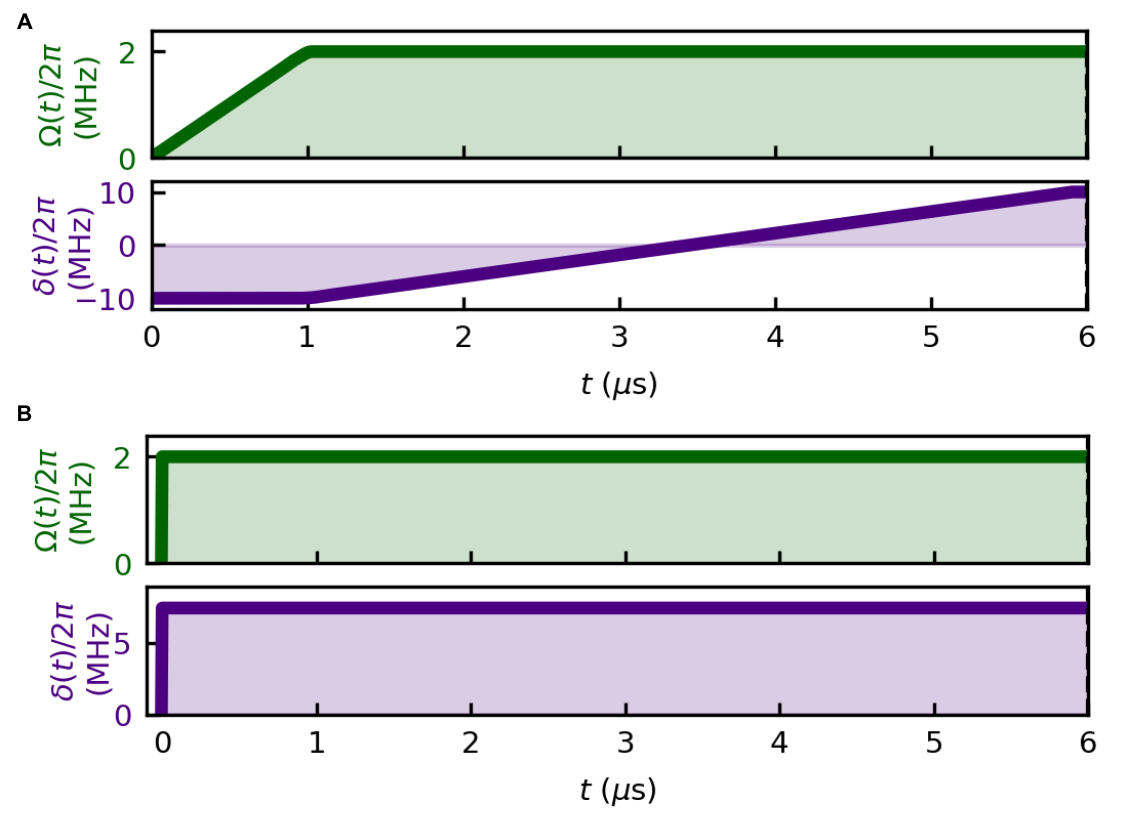}
     \caption{\textbf{Example of pulse sequences} \textbf{A,} used in the state-preparation protocols and \textbf{B,} in the post-quench dynamics protocols.}
     \label{fig:pulse_samples}
\end{figure}

\textbf{Programming protocols for adiabatic state preparation and post-quench dynamics}

Cloud-based access to the QPUs allows the user to tune the quantum simulator for various experiments, such as adiabatic state preparation or quenching the system and observing its response. We can adjust the geometry of the atomic register and the temporal profiles of the control pulses $\Omega(t)$ and $\delta(t)$ using \texttt{Sequence} objects of the \texttt{pulser} library~\cite{silverio_pulser_2022}. In this work, we use two families of pulse sequences $\Omega(t),\, \delta(t)$ to evolve the initial product state $\ket{\psi(t=0)}=\ket{g\,\dots\, g}$. 

On the one hand, to study the equilibrium properties of TmMgGaO$_4$ at low energies we use a quasi-adiabatic annealing schedule, depicted in Fig.~\ref{fig:pulse_samples}\textbf{A}: an initially large, but not infinite, negative detuning $\delta(t=0)=-3U_1/\hbar$ ($|\delta(t=0)|/2\pi=11.1$ MHz) brings $\ket{\psi(0)}$ close to being the ground state of $\hat{H}_{\text{QPU}}$ at $t=0$, up to a small admixture of unwanted states. The latter can manifest as oscillatory behaviour of observables, notably after the transition. We then use the AOM modulation to slowly ramp up $\Omega(t)$ to its maximum value $\Omega_{\rm max}/2\pi=2.0$ MHz required for matching the transverse field of the material, i.e. $\hbar\Omega_{\rm max}/U_1=1.08/2$ and $\delta(t)$ to the desired longitudinal field value between $\hbar\delta/U=-2$ $(\Delta_z/J_1\approx10)$ and $\hbar\delta/U_1=2.7$ $(\Delta_z/J_1\approx1)$. At each set point, the quantum state is approximately the ground-state of the target Hamiltonian, up to diabatic effects associated with a finite-time protocol ($T\sim5~\mu$s). To measure the quantum state, we perform a final step, consisting in ramping down $\Omega(t)$ to zero as fast as possible, to freeze the z-basis observables.

Finally, the atomic states are measured by switching the traps back and taking a site-resolved state-dependent fluorescence image where each state $\ket{g}$ ($\ket{r})$ is mapped to a bright (dark) spot. In this way, we sample 200-300 bitstrings per data point to estimate $z$-basis observables such as the local one-site magnetisation or two-site correlations. All cloud-related operations, including the submission of the \texttt{Sequence}, classical post-processing, and retrieval of the measurement results, occur on timescales on the order of seconds and are therefore negligible compared to the duration of the quantum simulation itself.

This quasi-adiabatic protocol was followed on the three devices FM1, FM2 and FC1 with slight variations. On FM1, the final ramp down of $\Omega(t)$ was performed with the EOM, and the traps were switched back right after, thus leaving minimal delay between the end of the \texttt{Sequence} and the measurement of the state. On FM2 and FC1, the ramp down was performed with the slower AOM, and the traps were switched off for a fixed duration $\Delta t = 6 \, \mu\mathrm{s}$, regardless of the \texttt{Sequence} duration. The cleaner implementation on FM1 led to reduced experimental noise, which is why we conducted most of the experiments on this device.

On the other hand, to study the non-equilibrium post-quench dynamics we use the square-pulse schedule depicted in Fig.~\ref{fig:pulse_samples}\textbf{B}: a fast ramp up of the QPU parameters to the material set point, which is kept constant during variable evolution times $100~\mathrm{ns} \leq t \leq 5.9~\mu\mathrm{s}$, and a fast ramp down before measurement. This protocol was carried out exclusively on FM1 in a similar manner than described above for quasi-adiabatic protocols. As the main difference, note that the EOM was used for both the ramp up and the ramp down of the control pulses. \\

\textbf{Magnetic observable estimators from QPU bitstrings}

\begin{figure*}
    \centering
    \includegraphics[width=1\linewidth]{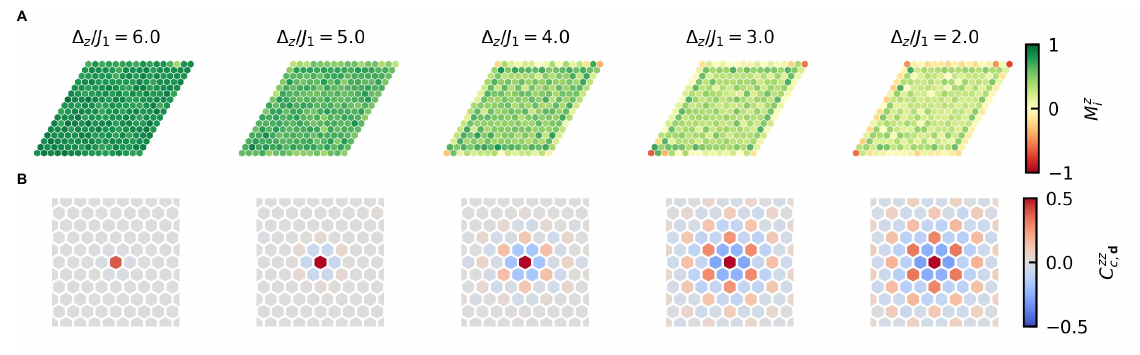}
    \caption{\textbf{Magnetisation and connected correlation patterns across the field-driven transition.}
    \textbf{A,} Site-resolved magnetisation $M^z_i$ measured on the QPU for a $N=256$ rhombus with decreasing longitudinal field $\Delta_z/J_1$, showing the evolution from a strongly polarised paramagnet to an ordered magnet with $1/3$ magnetisation.
    \textbf{B,} Corresponding connected correlation maps $S^{zz}_\text{QPU}(\mathbf{r}) = 
    1/N_{\mathbf{r} }\sum_i\langle\hat{\sigma}^z_i\hat{\sigma}^z_{i+\mathbf{r}}\rangle-\langle\hat{\sigma}^z_i\rangle\langle\hat{\sigma}^z_{i+\mathbf{r}}\rangle$ with $N_{\mathbf{r}}$ numbers of pairs separated by displacement vector $\mathbf{r}$.}
    \label{fig:obs-scans}
\end{figure*}

Local observables, such as magnetisation and correlation functions (Fig.~\ref{fig:obs-scans}\textbf{a-b}), are evaluated over the central 2D bulk region of size $N_\text{b}=(L-4)\times(L-4)$ within the full $N=L\times L$ array, corresponding to a two-site buffer from each edge to mitigate finite-size boundary effects. To further reduce perturbations arising from preparation defects, we retain only experimental shots that are free of defects within an additional one-site buffer surrounding the bulk region. 

False positive $\varepsilon$ and false negative $\varepsilon^\prime$ detection errors ~\cite{de_leseleuc_analysis_2018} are corrected at the level of single-body magnetisation and two-body correlation functions using independently measured error rates. The corrected magnetisation $M^z_i=\langle\hat\sigma^z_i\rangle$ per site is obtained from the measured value \(\tilde{M}^z_i\) as
\begin{equation}
M_i^z = (\tilde{M}^z_i - (\varepsilon^\prime - \varepsilon))/(1 - \varepsilon - \varepsilon^\prime)
\end{equation}
and similarly for corrected correlations $C^{zz}_{ij}=\langle\hat\sigma^z_i\hat\sigma^z_j\rangle$ from measured correlations $\tilde{C}^{zz}_{ij}$,
\begin{equation}
C^{zz}_{ij} = (\tilde{C}^{zz}_{ij} - (\varepsilon^\prime - \varepsilon)(\tilde{M}^z_i + \tilde{M}^z_j) - (\varepsilon^\prime - \varepsilon)^2)/(1 - \varepsilon - \varepsilon^\prime)^2
\end{equation}
For FM1, we measure $\varepsilon = 1\%$ and $\varepsilon^\prime = 4\%$ while for FM2, FC1, Jade and Ruby we measure $\varepsilon = 1\%$ and $\varepsilon^\prime = 7\%$.

Although bitstring filtering and correction recover signals closer to those of an ideal, noiseless QPU, our conclusions are robust to this processing.  
Direct evaluation of observables from raw bitstrings yields quantitatively consistent results.  
Detection errors only introduce vertical offsets or rescalings and do not affect, for example, the estimate of the quantum critical point.

\vspace{15pt}
\textbf{QPU estimation of the quantum critical point}

\begin{figure}
    \centering
     \includegraphics[width=0.5\linewidth]{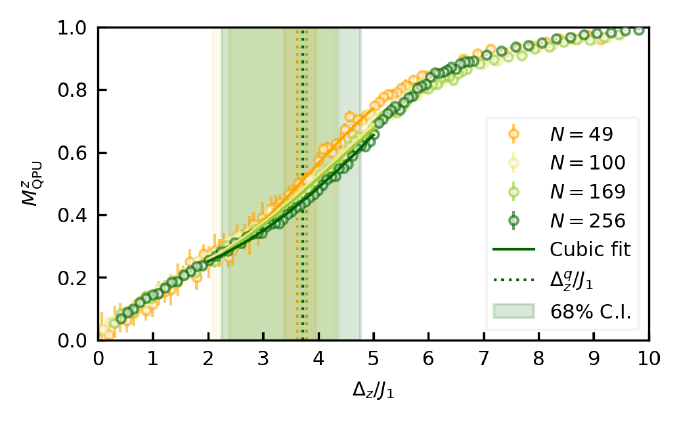}
     \caption{\textbf{Quantum critical point estimation from QPU measurements.} Magnetisation $M^z_{\rm QPU}$ as a function of the longitudinal field $\Delta_z/J_1$ for system sizes $N = 49, 100, 169, 256$ (dots). The estimates of the quantum critical point $\Delta_z^q/J_1(N)$ (dashed) are obtained by computing the maximum of a cubic polynomial fit (solid) performed over a finite window around the expected transition; the shaded regions indicate the 68\% confidence interval obtained by varying the fit window.}
     \label{fig:scaling}
\end{figure}

To estimate the quantum critical point with the QPU measurements, we access the raw signals $M^z_{\rm QPU}$ (Fig.~\ref{fig:scaling}) and $S^{zz}_{\rm QPU}(\mathbf{q_{1/3}})$ (Fig.~\ref{fig:fig2}\textbf{D}) for various system sizes $N = 49, 100, 169, 256$ .
In principle, the quantum critical point can be identified where the observables change most rapidly: $M^z_{\rm QPU}$ exhibits a sharp drop and $S^{zz}_\text{QPU}(\mathbf{q_{1/3}})$ a pronounced increase, corresponding to the maximum of their derivatives with respect to $\Delta_z/J_1$. In practice, the QPU measurements are affected by shot-noise-induced fluctuations, which can obscure the precise location of the transition. To mitigate this effect, we perform a cubic polynomial fit over a finite window around the expected phase transition, typically in the range $2\leq\Delta_z/J_1\leq6$. The maximum of the derivative of the smooth fit provides a robust estimate of the quantum critical point while suppressing shot-noise effects. To reduce sensitivity to the fitting window, we repeat the procedure over multiple overlapping windows of comparable length and take the mean value as the final estimator. For the magnetisation (Fig.~\ref{fig:scaling}), the weak curvature of the signal leads to a strong dependence of the fit on the window choice, resulting in large confidence intervals. In Fig.~\ref{fig:fig2}\textbf{D}, we present the extracted critical points for $S^{zz}_\text{QPU}(\mathbf{q})$ only for the largest systems ($N=256$), for which bulk behaviour dominates and the growth of the order parameter is significant. We obtain $\Delta_z^q/J_1(N=256) = 3.87_{-0.36}^{+0.44}$, where the uncertainty corresponds to the $68\%$ confidence interval derived from a varying fit windows method.\\

\section*{Supplementary Note 3: Theoretical mapping}\label{note:mapping}

\textbf{Mapping the Rydberg QPU and TmMgGaO$_4$ through the triangular lattice Ising model}

In a triangular lattice geometry, and using only Pauli matrices, the QPU Hamiltonian of Eq.~\eqref{eq:ham_qpu_1} can be rewritten in terms of Eq.~\eqref{eq:ising_tmgo}, neglecting the global change in energy scale ($\hat{H}_\text{QPU}\approx \alpha_\text{QPU}\hat{H}_\text{TMGO}$), as
\begin{equation}\label{eq:ham_qpu_2}
\begin{split}
    & \hat{H}_{\text{QPU}}=\hat{H}_\text{TMGO}+\hat{H}_\text{diff.}.
\end{split}
\end{equation}
Here we identified $\hbar J_1=U_1/4=C_6/(4r_1^6)$ as the nearest-neighbour interaction, $r_1$ being the lattice spacing. We also identify $\Delta_x(t) = \frac{\Omega(t)}{2}$, and $\Delta_z(t)=\frac{1}{2}\left[\delta_U-\delta(t)\right]$, with $\hbar\delta_U=\frac{1}{2}\sum_{ij}U_{ij}/N$ and $U_{ij}=C_6/r_{ij}^6$. The last term $\hat{H}_\text{diff.}$ accounts for the difference between $\hat{H}_{\text{QPU}}$ and $\hat{H}_{\text{TMGO}}$,
\begin{equation}
\begin{split}
    \frac{\hat{H}_{\text{diff.}}}{\hbar} = &\sum_i\Delta_{z,i}\hat \sigma^z_i +  \sum_{\langle i,j\rangle_{n>2}} \frac{U_{ij}}{4\hbar}\hat \sigma^z_i\hat \sigma^z_j
    -\frac{1.3J_1}{100}\sum_{\langle i,j\rangle_2}\hat \sigma^z_i\hat \sigma^z_j.
\end{split}
\end{equation}
On the one hand, $\hbar\Delta_{z,i}=\sum_{l,j}U_{lj}/(4N)-\sum_{j}U_{ij}/4$ is a site-dependent longitudinal field, which is non-zero for site-dependent interaction profiles $\sum_{j}U_{ij}$. This interaction profile is constant in the bulk of the triangular lattice, so it has a negligible impact in the large system size limit, where variations of the interactions at the edges are not expected to play a role.
On the other hand, the last two terms account for, respectively, the tail of the $r_{ij}^{-6}$ Rydberg interactions beyond next-nearest-neighbours (e.g., $\langle i,j\rangle_3$ corresponds to third-order neighbours), and the small difference in the next-to-next-nearest interaction $J_2$. Such small corrections are not expected to change the properties of the model, but can slightly modify the parameter regimes. 

This setup naturally  enables faithful simulation of the two-dimensional quantum Ising Hamiltonian in Eq.~\eqref{eq:ising_tmgo}. The effective longitudinal field is controlled via the laser detuning as $\Delta_z/J_1 = \hbar [\delta_U - \delta(t)]/(2 U_{\rm 1})$. The transverse field is controlled by the Rabi frequency as $\Delta_x/J_1 = 2 \hbar \Omega / U_{\rm 1}$.\\

\section*{Supplementary Note 4: Material Synthesis and Characterization}\label{note:material}
\textbf{Crystal Growth}
 
\begin{figure*}[h]
    \centering
     \includegraphics[width=0.4\linewidth]{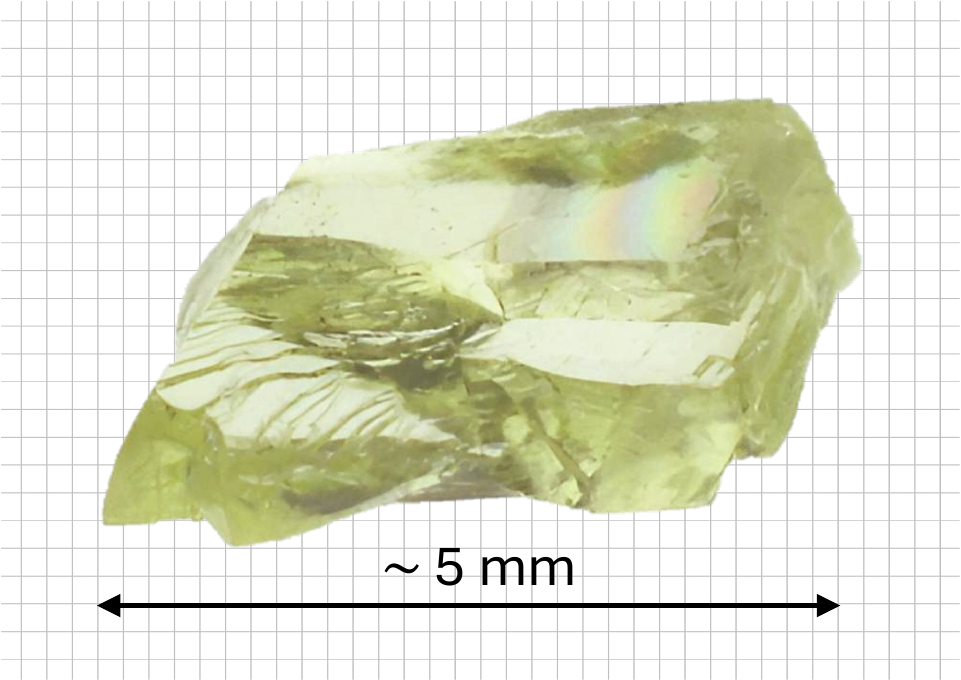}
     \caption{\textbf{Macroscopic photograph of the TmMgGaO$_4$ single-crystal sample synthesised.} The exact colours of the sample in this photograph are influenced by lighting conditions and should not be relied upon for precise colour representation. The samples appear white and slightly opaque under normal viewing conditions.}
     \label{fig:crystal-picture}
\end{figure*}

 A polycrystalline sample of TmMgGaO$_{4}$ shown in Fig.~\ref{fig:crystal-picture} was synthesised by solid state reaction. Stoichiometric ratios of Tm$_{2}$O$_{3}$, MgO and Ga$_{2}$O$_{3}$ fine powder were ground and reacted at a temperature of 1450 Celsius degree for 3 days with several intermediate grindings. The single-crystal sample of TmMgGaO$_{4}$ was grown using the optical floating-zone method under a 5 atm oxygen atmosphere. The best single crystal was obtained with a pulling speed of 1.0 mm/per hour. The obtained transparent crystals exhibit cleavable [001] surface. The crystal structure consists of triangular layers of TmO$_6$ octahedra separated by non-magnetic (Mg,Ga)O$_5$ layers. \\

 \textbf{Magnetic characterisation of crystals}
 
 \begin{figure*}
    \centering
    \includegraphics[width=\linewidth]{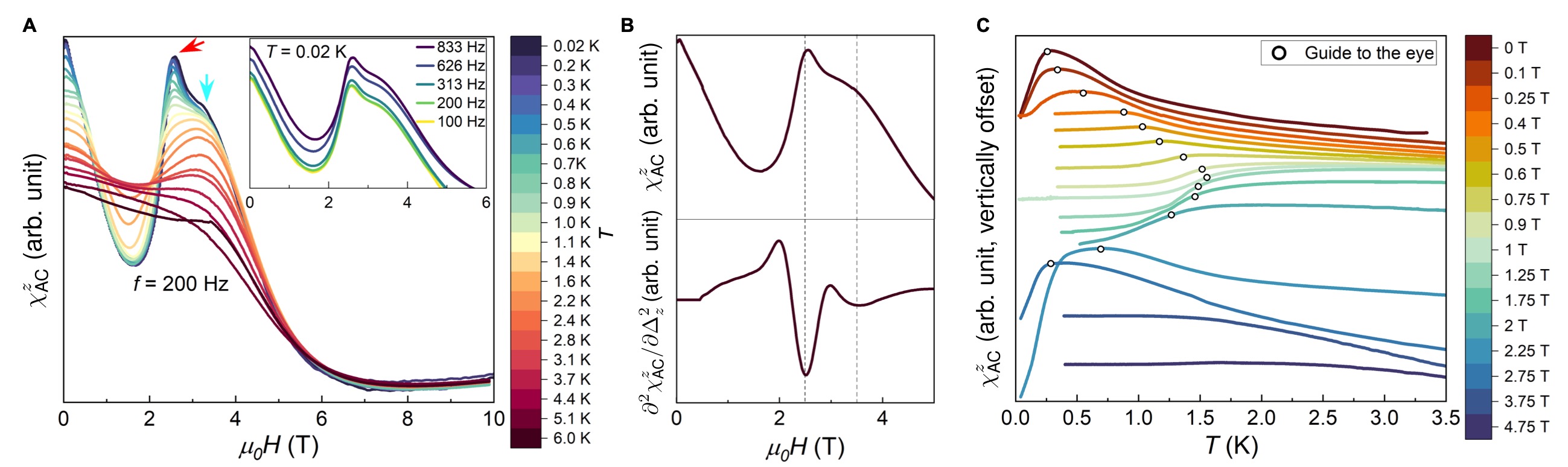}
    \caption{\textbf{Estimation of quantum and thermal critical points through AC susceptibility measurements of TmMgGaO$_4$.} \textbf{A,} AC susceptibility as a function of the applied magnetic field along the c axis at different temperatures and for excitation frequency of $200$ Hz Inset: Frequency dependence of $\chi^z_{\mathrm{AC}}$ at $0.02$ K. \textbf{B,} Comparison of $\chi^z_\text{AC}(20\text{mK})$ and its second field derivative. The minima in the second derivative are used to locate the $\mu_0 H$ fields corresponding to the quantum phase transition and the thermal bump in Fig.~\ref{fig:fig2}\textbf{B}. The derivatives were smoothed using a Savitzky–Golay filter with a 50-point window. \textbf{C,} Temperature dependence of $\chi^z_{\mathrm{AC}}$ for different magnetic fields applied along the $c$ axis. The curves are offset in the y-axis for better visualisation. The white points serve as guides to the eye, indicating the positions extracted for the thermal dome (T sweeps) in Fig.~\ref{fig:fig2}\textbf{B}.}
    \label{fig:chi_ac}
\end{figure*}
 
 AC susceptibility $\chi^z_{\rm AC}$ measurements presented in Fig.~\ref{fig:fig2}\textbf{A} are extended in Fig.~\ref{fig:chi_ac}\textbf{A}. They were conducted at the National High Magnetic Field Laboratory (NHMFL) using a conventional mutual-inductance technique with a custom-built susceptometer~\cite{dun_chemical_2014}. Experiments were performed in a superconducting magnet providing fields up to 18~T and over a temperature range from 20~mK to 6~K. An AC excitation field of 0.83 Oe and frequency of 200~Hz. No measurable frequency dependence is observed at excitation frequencies of a few hundred hertz as shown in the inset of Fig.~\ref{fig:chi_ac}\textbf{A}, indicating that the system remains in quasi-equilibrium on the timescale of the measurement. 

Complementary DC measurements from $T=5$K to 2K are discussed in the next paragraph.
 
 Under these conditions, the AC susceptibility coincides with the equilibrium DC susceptibility $\chi^z_{\rm DC}$ and the magnetisation can be reconstructed from the AC susceptibility by field integration. In particular, the spin-1/2 magnetisation $M^z$ defined for Eq.~\eqref{eq:ising_tmgo}, and shown in Fig.~\ref{fig:fig1}\textbf{E}, is expressed as 
\begin{equation}
    M^z_\text{AC}(\mu_0 H)\equiv \frac{\int_{0}^{\mu_0 H}\chi^z_{\text{AC}}(\mu_0 H')\,\text{d}\mu_0 H'}{\int_{0}^{10 \text{T}}\chi^z_{\text{AC}}(\mu_0 H')\,\text{d}\mu_0 H'},
\end{equation}
where we use the physical normalisation $M^z_\text{AC}(0)=0$ and $\lim_{\mu_0 H\to \infty} M^z_\text{AC}(\mu_0 H)\approx M^z_\text{AC}(10\,\text{T}) =1$.\\

\textbf{Additional DC magnetisation measurements}\label{si-sec:dc-meas}

\begin{figure*}[h]
    \centering
     \includegraphics[width=\linewidth]{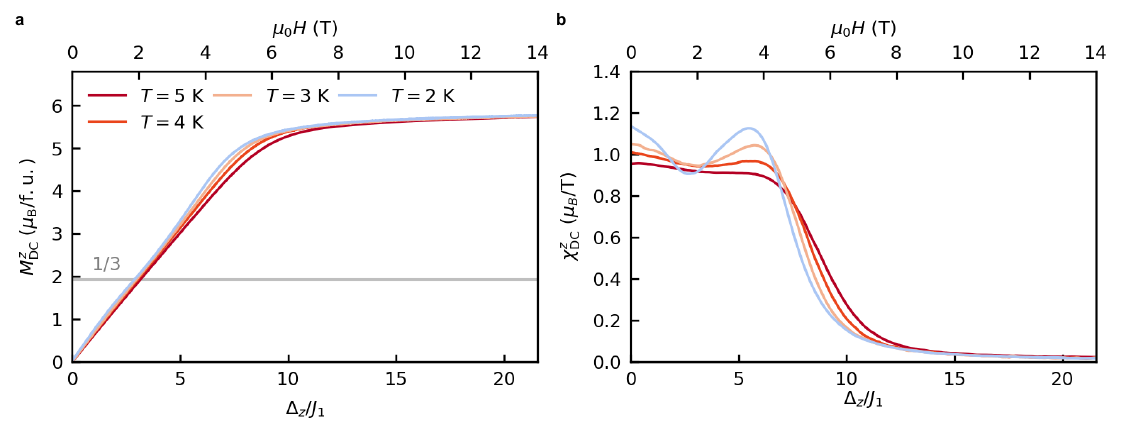}
     \caption{\textbf{Measurements of DC magnetisation of TmMgGaO$_4$.} \textbf{A,} DC magnetisation $M^z_{\mathrm{DC}}(\Delta_z/J_1)$ measured at decreasing $T=5,4,3$ and $2$K; \textbf{B,} differential DC susceptibility $\chi^z_{\rm DC}$. }
     \label{fig:DC}
\end{figure*}

As a cross-check, we also performed DC magnetisation measurements on single-crystal TmMgGaO$_4$ samples using vibrating sample magnetometry (VSM) in a 14~T Quantum Design Physical Property Measurement System (PPMS). The magnetic field was applied along the magnetic easy $c$ axis. Single crystals with typical masses of $\sim$3~mg were mounted on quartz sample holders using General Electric varnish. Measurements were carried out at temperatures down to 2~K. DC magnetisation $M^z_{\rm DC}$ and differential susceptibility $\chi^z_{\rm DC} = \mathrm{d}M^z_{\rm DC}/\mathrm{d}H$, measured between $5\,\mathrm{K}$ and $2\,\mathrm{K}$ (Fig.~\ref{fig:DC}), exhibit an elbow (or quasi-plateau) around $\Delta_z/J_1 = 4$, consistent with the $1/3$ phase. In this temperature range, however, $\chi^z_{\rm DC}$ shows no distinct peak associated with this feature.

\section*{Numerical Methods and Noise Model}\label{note:numerics}

\textbf{TDVP numerical simulations} 

\begin{figure*}
    \centering
    \includegraphics[width=\linewidth]{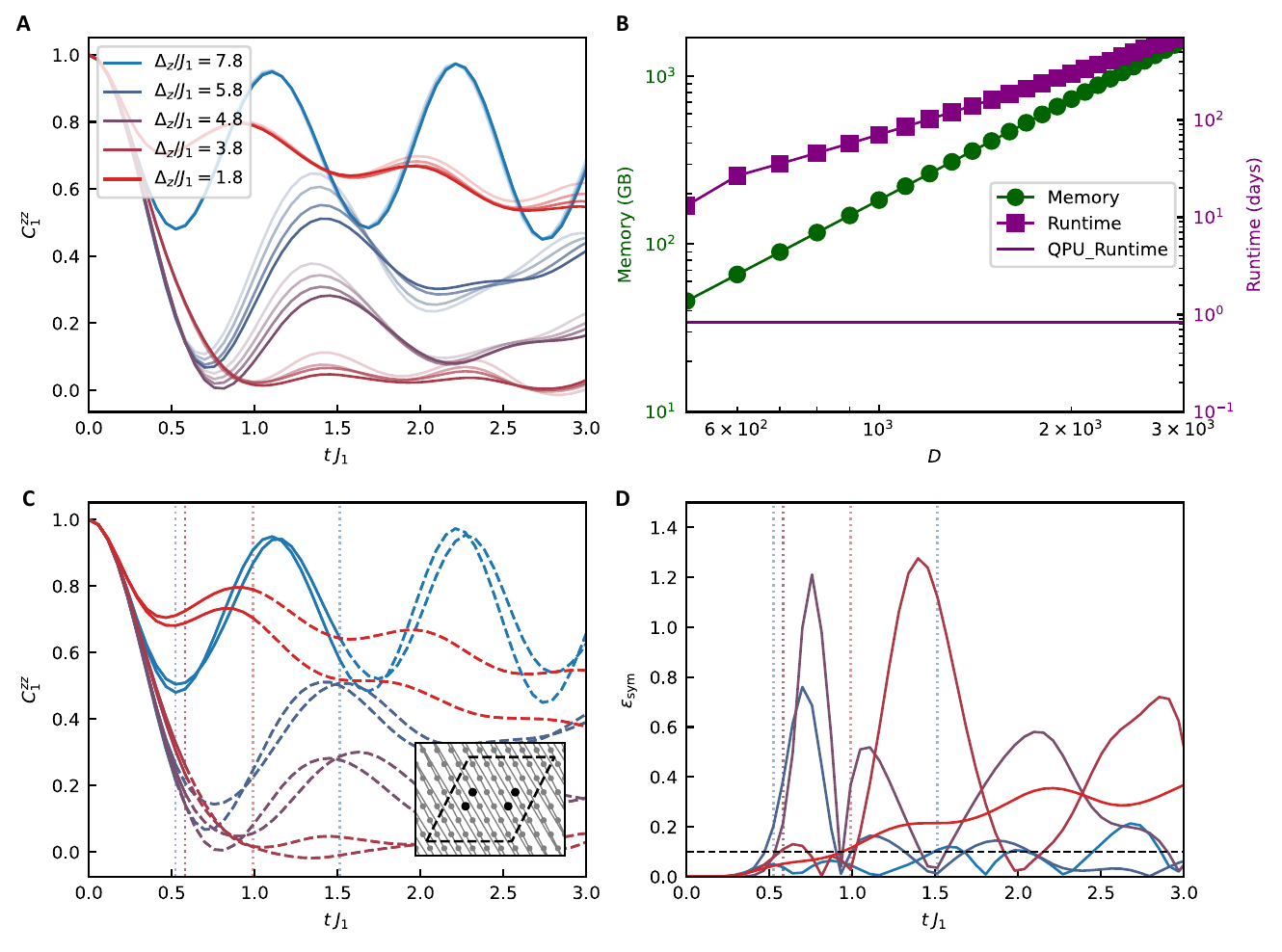}
    \caption{\textbf{Breakdown of MPS convergence in post-quench dynamics.} 
    \textbf{A,} Nearest-neighbour correlations, $C_{1}^{zz}$, following an abrupt quench of the magnetic field, computed using MPS with increasing bond dimension $D$. For each value of $\Delta_z/J_1$, the most translucent curve corresponds to the lowest bond dimension $D=75$, with $D$ subsequently doubled up to $D=600$. \textbf{B,} Estimated computational resources required to increase the bond dimension, see \SRefSec{si-sec:resource estimation} for details. QPU runtime is also included for comparison. \textbf{C,} NN correlations at the largest bond dimension considered ($D=600$), highlighting pronounced asymmetries between symmetry-equivalent NN pairs in the bulk. The inset illustrates the snake-like MPS mapping of the two-dimensional lattice and marks the two symmetry-equivalent NN pairs within the $6\times6$ bulk used in the comparison. \textbf{D,} Relative asymmetry error $\varepsilon_{\text{sym}}$ between the two NN pairs as a function of time. Vertical dotted lines indicate the times at which the relative error first exceeds a threshold of $10\%$.}\label{fig:resource_estimates}
\end{figure*}

The MPS simulations presented in this work are performed using the Python package \texttt{emu-mps}~\cite{bidzhiev_efficient_2025}, a tensor-network backend of the Python package \texttt{pulser}~\cite{silverio_pulser_2022}.
Time evolution is implemented via the two-site TDVP algorithm with an MPS ansatz. We define the MPS via snaking path across a triangular rhombus lattice of linear size $N=L \times L$.
The path starts from one corner of the lattice and proceeds along one lattice row before repeating the next row, thereby covering all sites in a one-dimensional ordering compatible with the lattice geometry. Furthermore, the time step for the evolution is set to \(\Delta t \sim 10^{-3} J_1^{-1}\), ensuring sufficient temporal resolution of the dynamics. These simulations are carried out using the largest available bond dimension for each lattice size, chosen in multiples of 100, and are executed on an NVIDIA A100 GPU with 40~GB. For the most computationally challenging simulations, namely post-quench dynamics in large systems, we perform this snaking diagonally to best preserve locality with the MPS mapping and the hardware was upgraded to an H200 NVIDIA GPU with 141~GB of memory. In order to gauge the accuracy of the all MPS simulations we consider both convergence with respect to increasing bond dimensions, see Fig.~\ref{fig:mps-scaling}, as well as the preservation of geometric symmetries of the Hamiltonian, see Fig.~\ref{fig:resource_estimates}. Full details of the convergence analysis, symmetry diagnostics and resource scaling are provided, as well as a discussion on other state-of-the-art numerical methods, in \SRefSec{si-sec:resource estimation}.\\

\begin{figure*}
    \centering
    \includegraphics[width=\linewidth]{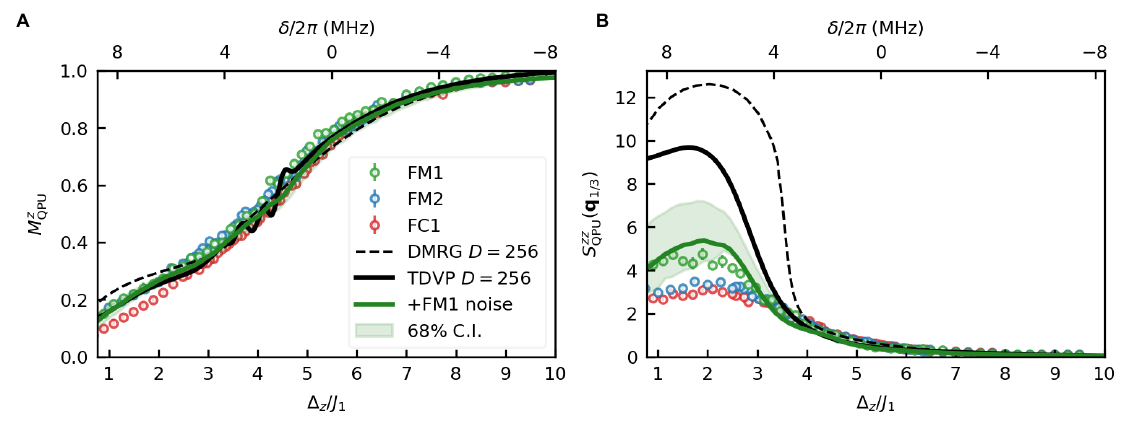}
    \caption{\textbf{Comparison between DMRG, TDVP, noisy TDVP, and various QPU observables.}
    \textbf{A,} $M^z_{\rm QPU}$ and \textbf{B,} $S^{zz}_{\rm QPU}(\mathbf{q}_{1/3})$ as functions of the applied field $\Delta_z/J_1$ for $N=100$.
    Experimental measurements (with error bars reflecting shot noise), obtained on three devices---FC1, FM2, and FM1 (red, blue, and green symbols, respectively)---are compared with numerical results from DMRG (dashed lines), noiseless TDVP, whose convergence is discussed in \SRefSec{si-sec:resource estimation} (solid black line), and noisy TDVP simulations.
    The latter include the Monte Carlo mean (solid green line) and standard deviation (shaded region) over 100 trajectories designed to mimic FM1 hardware noise (see Supp. Mat.). Compared to the ground-state calculation (DMRG), the finite-time simulation (TDVP) gives a slightly shifted magnetisation curve with diabatic oscillations after crossing $\Delta^z_z\approx 6$, and a reduction of the maximum structure factor ($\approx 20\%$). Noisy simulations are compatible with the data obtained in the finite-time protocol of FM1: the diabatic oscillations of the magnetization curve are smoothened, and the maximum structure factor is further reduced by approximately $50\%$. Nevertheless, the noise levels remain sufficiently low to clearly resolve the distinct signatures of the two phases studied in this work. Slightly higher noise levels in FM2 and FC1 naturally account for the additional discrepancies observed relative to FM1. 
    }
    \label{fig:noise_study}
\end{figure*}

\textbf{Phase diagram of periodic system using DMRG}

The phase diagram $(\Delta_z/J_1,\Delta_x/J_1)$ of the transverse field Ising model on the triangular lattice presented in Fig.~\ref{fig:fig1}\textbf{D} is computed using the density matrix renormalisation group (DMRG) algorithm implemented in \texttt{TeNPy} library (version 1.0.6) \cite{tenpy2024}. Numerics are performed on a rhombic cluster of size $N=6\times6$ with periodic boundary conditions. Ground and low-lying excited states are obtained via variational optimisation of MPS with a fixed bond dimension $D=256$, taken to be sufficient for constructing a qualitative sketch of the phase diagram. Quasi-degenerate states are accessed by rerunning DMRG on Hamiltonians truncated by projecting out previously obtained states. These states are sampled in the computational basis, and their bitstrings are combined with equal weights to compute the structure factor $S^{zz}_{\rm QPU}(\mathbf{q}_{1/3})$, renormalised between $0$ and $1$ (dividing by $(2/3)^2N$) and used to distinguish the paramagnetic and $1/3$-ordered phases. Owing to the finite system size, quantum critical points are shifted relative to their thermodynamic-limit values, reflecting the rounding of phase transitions and the limited resolution of long-wavelength correlations on this small finite cluster. 
We use the same method for the DMRG curves of Fig.~\ref{fig:noise_study} for a $10\times 10$ triangular lattice at bond dimension $D=256$.\\
\textbf{TDVP numerical simulations of QPU noise}

On the timescales relevant to the few-microsecond protocols explored in this work, the QPU dynamics is well described by coherent unitary evolution under the effective Hamiltonian of Eq.~\eqref{eq:ham_qpu_1}. The experimental implementation is nevertheless subject to several sources of noise that affect both the Hamiltonian parameters and the measured observables~\cite{scholl_quantum_2021,scholl_quantum_2021-1,de_leseleuc_analysis_2018,noisemodel_inprep}. Here we summarise the dominant contributions, characterised independently on the FM1 device.

Finite atomic temperature induces residual atomic motion in the traps and shot-to-shot position fluctuations, leading to variations in interaction strengths. Additional fluctuations arise from Doppler shifts associated with site-dependent atomic velocities, leading to variations of local detunings. Technical noise in the driving lasers further introduces shot-to-shot and time-dependent fluctuations of the Rabi frequency and detuning, including components that cause in-sequence dephasing. Imperfections in the pulse-shaping hardware, arising from finite bandwidth, result in smooth pulse edges and distortions relative to the target control waveforms. Spatial inhomogeneities due to the finite waist of the addressing beams lead to position-dependent variations of the control fields.

Decoherence mechanisms outside the ideal qubit manifold include spontaneous decay and dephasing, originating from finite Rydberg lifetimes, black-body-radiation-induced processes, and spontaneous emission to the intermediate state of the three-level transition during excitation. Imperfect state preparation leads to a small fraction of atoms being initialised outside the qubit manifold, and thus not participating in the many-body dynamics. Finally, state readout is affected by detection errors, including both false positives, when atoms in $\ket{g}$ are not recaptured at the end of the protocols and false negatives, when a decay happens during the imaging process for instance.

These effects are simulated using the \texttt{pulser} library~\cite{silverio_pulser_2022} and the TDVP tools of \texttt{emu-mps}~\cite{bidzhiev_efficient_2025}. For the FM1 device, expectation values are averaged over up to $100$ noisy Hamiltonian realisations, decay processes are included via a quantum jump formalism, and detection errors are applied by sampling each final state $200$ times. In Fig.~\ref{fig:noise_study} we use this method to characterize the noise impact for the quantum phase transition scan at $N=100$ presented in Fig.~\ref{fig:fig2}.\\


\begin{figure*}
    \centering
    \includegraphics[width=1.0\linewidth]{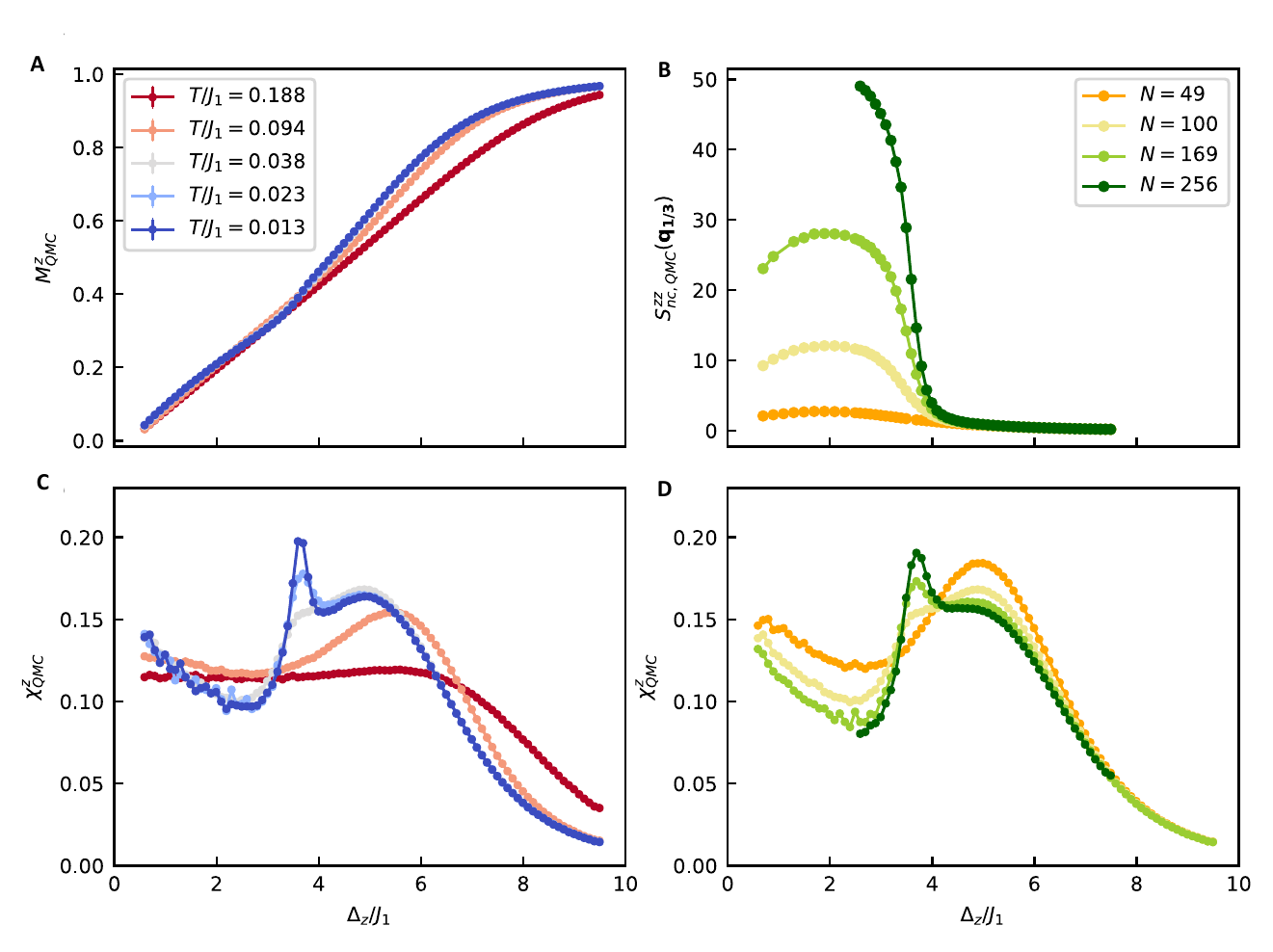}
    \caption{{\bf Quantum Monte Carlo study of the quantum phase transition in the Rydberg Hamiltonian.} \textbf{a (c)}, average local magnetisation, $M_{\mathrm{QMC}}=\frac{1}{N}\sum_{i=1}^N\langle\sigma_i^z\rangle$ (susceptibility $\chi_\text{QMC}=\partial M_{\mathrm{QMC}}/\partial \Delta_z$) as a function of the longitudinal field $\Delta_z/J_1$ for $N=100$ and various normalised temperatures $T/J_1$. Arrows highlight the emergence of low-temperature features in the simulations, consistent with the temperature dependence in Fig.~\ref{fig:fig2}\textbf{A}. \textbf{b (d),} size dependence of the non-connected structure factor, $~S^{zz}_{nc,\mathrm{QMC}} = 1/{N_b} \sum_{i,j \in \mathrm{bulk}} e^{i \mathbf{q} \cdot \mathbf{r}_{ij}} \langle \hat{\sigma}_i^z \hat{\sigma}_j^z \rangle$ (susceptibility $\chi_\text{QMC}$), for $T / J_1 \approx 0.038$, corresponding to $T \approx 0.108\,\mathrm{K}$. The slow increase of $S^{zz}_{nc,\mathrm{QMC}}$ at the transition point and the absence of a clear peak in $\chi_\text{QMC}$ for $N<169$ highlights the need for larger system sizes in QPU aiming at detecting the transition.}

       \label{fig:QMC_data}
\end{figure*}

\textbf{Quantum Monte Carlo simulations of the Rydberg Hamiltonian}

We use a quantum Monte Carlo approach based on the finite-temperature stochastic series expansion (QMC-SSE) algorithm \cite{sandvik_sthocastic_2003,merali_stochastic_2024} to simulate thermal equilibrium states of the Rydberg Hamiltonian $\hat{H}_\textrm{QPU}$ [Eq.~\eqref{eq:ham_qpu_1}]. This algorithm relies on a classical representation obtained from a Taylor expansion of the partition function and is implemented using both local and cluster updates. 
Consistently with the QPU setup, we perform simulations of a triangular atomic array with a rhombus geometry, containing $N = L \times L$ qubits. In all simulations, we use $N_{\mathrm{therm}} = 5 \times 10^4$ Monte Carlo steps for thermalisation, and observables are evaluated using $N_{\mathrm{meas}} = 10^7$ successive measurements. A binning analysis is then used to estimate errors.

In Fig.~\ref{fig:QMC_data} we use this method to benchmark the effect of temperature and finite size in the paramagnet to $1/3$ order quantum phase transition studied through magnetic and QPU measurements in Fig.\ref{fig:fig2} (Fig.~\ref{fig:QMC_data}). Furthermore, we also use this method for the characterisation of the long-time QPU dynamics described in the next section.\\

\textbf{Effective temperature of post-quench dynamics}

We probe whether the post-quench dynamics of quantum simulations performed on the QPU
leads to thermalisation in the long-time regime in two steps.

First, we compute the effective temperature associated with the energy of the initial state
$ \ket{\psi(0)} = \ket{\uparrow \dots \uparrow}$.
In particular, we assume energy conservation and match the initial energy with that of a
thermal ensemble at an effective temperature $T$,
\begin{equation}
\bra{\psi(0)} \hat{H}_{\mathrm{QPU}} \ket{\psi(0)} =
\frac{Tr\left( \hat{H}_{\mathrm{QPU}} e^{- \hat{H}_{\mathrm{QPU}}/(k_\text{B}T)} \right)}{Z},
\end{equation}
where $\hat{H}_{\mathrm{QPU}}$ is defined in Eq.~\eqref{eq:ham_qpu_1} and $Z$ is its partition function at temperature $T$.
We use the QMC-SSE simulations described in the previous section to compute $k_\text{B}T/U_1$ [or $k_\text{B}T/(\hbar J_1)$] for different values of $\hbar\delta/U_1$ (or $\Delta_z/J_1$). In practice, this is achieved by numerically computing the temperature dependence of the thermal
energy and matching it to the energy of the initial state. The results for different values of $\Delta_z/J_1$ are shown in Fig.~\ref{fig:fig.4}{\bf c}.
Interestingly, we observe that quenches to parameter regimes of the ground-state phase diagram associated with the $1/3$ phase lead to an effective negative temperature. For example, for $\Delta_z/J_1 = 1.8$, we obtain $k_\text{B}T/(\hbar J_1) = -1.25$.

As a second step to probe thermalisation, we compute observables at $T$ using the QMC-SSE approach and compare their values with the results obtained on the QPU as shown in Fig.~\ref{fig:fig.4}{\bf c}.

\textbf{Convergence and resource estimation for TDVP methods}\label{si-sec:resource estimation}

In this section we assess the performance and limitations of MPS simulations for the dynamics considered in this work. We start by discussing the convergence of the TDVP simulations of the quasi-adiabatic protocol used for ground-state preparation across the quantum phase transition at $N=100$ (Fig.~\ref{fig:noise_study}). Fig.~\ref{fig:mps-scaling} shows the bond-dimension dependence of two observables: the magnetization $M^z_{\rm QPU}(\Delta_z/J_1)$ in panel~\textbf{A} and structure factor $S^{zz}_{\rm QPU}(\mathbf{q}_{1/3})$ in panel~\textbf{B}. For both observables we show simulation results obtained with increasing bond dimensions, starting from $D=128$, doubling up to $D=512$, and finally $D=700$, as larger bond dimensions (e.g.\ $D=1028$) could not be accommodated within the 40~GB GPU memory. The insets of both panels show the relative difference between the simulations of a given bond dimension and the $D=700$ results. As the bond dimension is increased, the mean magnetisation exhibits clear convergence, indicating that the bond dimensions considered here are sufficient to accurately capture this observable.
Increasing the bond dimension modifies the short-time diabatic oscillations of $M^z_{\rm QPU}$ and slightly shifts the magnetisation in the $1/3$ plateau region toward lower values. In contrast, convergence is less evident for the structure factor. Nevertheless, a systematic decrease of $S^{zz}_{\rm QPU}(\mathbf{q}_{1/3})$ with increasing bond dimension (about $10\%$ between $D=256$ and $700$) is clearly visible at low $\Delta_z/J_1$, highlighting its greater sensitivity to residual truncation effects.
Although TDVP at $D=700$ remains computationally demanding and still not fully converged, these trends are essential for interpreting comparisons with experimental data. For this reason, and for computational tractability, all noisy TDVP simulations presented in Fig.~\ref{fig:noise_study} are performed at $D=256$.

We now turn to the discussion of the post-quench dynamics TDVP simulations, presented in Fig.~\ref{fig:fig.4} of the main text. Fig.~\ref{fig:resource_estimates} highlights the rapid breakdown of MPS simulations for post-quench dynamics in large two-dimensional systems, even at the largest bond dimensions, $D$, that are accessible with a H200 141GB GPU. In particular Fig.~\ref{fig:resource_estimates}\textbf{A} shows the nearest-neighbour correlation function $C^{zz}_1$ of an atomic pair in the bulk following an abrupt quench of the transverse magnetic field, computed for several values of $\Delta_z/J_1$ and increasing bond dimension. Despite systematically doubling the bond dimension from $D=75$ up to $D=600$, the resulting dynamics remain strongly dependent on $D$ across essentially the entire time window shown for most values of $\Delta_z/J_1$. Fig.~\ref{fig:resource_estimates}\textbf{B} provides estimates of the computational resources required to simulate the full long-time dynamics presented in Fig.~\ref{fig:fig.4} for increasing bond dimensions beyond the values accessible in our simulations. These estimates are obtained by fitting the theoretical scaling laws of TDVP to numerical benchmark data, see Ref.~\cite{vovrosh_resource_2025} for details. Both the projected memory footprint and wall-clock runtime grow rapidly with $D$, rendering such extensions prohibitive on currently available hardware. For reference, we also include a comparison line corresponding to the QPU runtime, assuming measurements taken every $\delta t J_1 = 0.1$, with 300 shots per point and a shot repetition rate of $0.7,\mathrm{Hz}$, as inferred from the experimental data.

Beyond the absence of quantitative convergence, the MPS results exhibit clear qualitative pathologies.
As shown in Fig.~\ref{fig:resource_estimates}\textbf{C}, even at $D=600$ the NN correlations associated with symmetry-equivalent bonds in the bulk, shown in the inset, differ significantly from one another, even when bond dimension scaling would lead you to believe the results have converged. Such asymmetries are unphysical and arise from the effective one-dimensional ordering imposed by the MPS snake mapping of the two-dimensional lattice, combined with insufficient entanglement capacity. To quantify this effect, in Fig.~\ref{fig:resource_estimates}\textbf{D} we introduce a relative asymmetry error,
\begin{equation}
    \varepsilon_{\text{sym}} = 2\frac{|C_1^{zz}(i)-C_1^{zz}(j)|}{\max(|C_1^{zz}(i)+C_1^{zz}(j)|,0.05)},
\end{equation} 
between two symmetry-related NN pairs, here indexed by $i$ and $j$, see \cite{vovrosh_simulating_2025} for further details. For all parameter regimes considered, $\varepsilon_{\text{sym}}$ grows rapidly in time and exceeds a conservative threshold of $10\%$ well before the end of the simulated evolution, as indicated by the vertical markers. This symmetry-based diagnostic provides a stringent and physically motivated criterion for MPS reliability, demonstrating that truncation-error-based measures alone can substantially overestimate the accessible simulation time in two dimensions. In Fig.~\ref{fig:fig.4} of the main text, the largest relative asymmetry of a given symmetrically equivalent group of NN pairs is considered for the convergence criteria of the bulk-averaged data.

Finally, we emphasize that the choice of MPS in this work is motivated by versatility rather than optimal accuracy. In recent work~\cite{vovrosh_simulating_2025}, we performed a systematic comparison of several state-of-the-art classical approaches for two-dimensional quantum dynamics, including neural quantum states (NQS), two-dimensional tensor networks (2DTN), tree tensor networks (TTN), and MPS. While MPS was not always the most accurate method across all regimes, it consistently proved to be the most flexible and robust across a wide range of dynamical protocols. In particular, NQS were found to struggle with quench dynamics of the type considered here. TTN approaches did not accurately reproduce the dynamics of correlation functions, even in regimes where local observables appeared well behaved. Moreover, both the triangular lattice geometry and the long-range nature of the interactions introduce additional challenges for current 2DTN implementations, which are the subject of ongoing and future work. For these reasons, MPS provides a natural and well-controlled baseline for diagnosing the limitations of classical simulations in the parameter regimes explored in this study.

\begin{figure*}
    \centering
    \includegraphics[width=\linewidth]{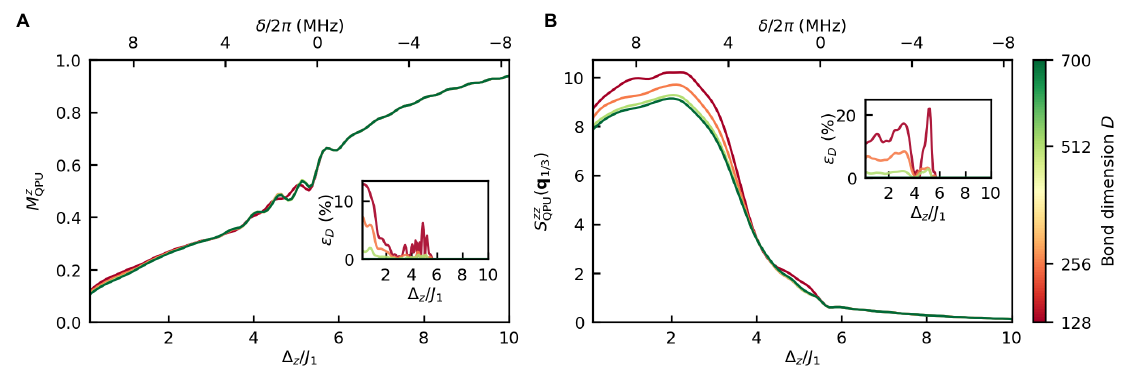}
    \caption{\textbf{Influence of bond dimension $D$ on TDVP observables.} \textbf{A,} Mean magnetisation $M^z_{\rm QPU}(\Delta_z/J_1)$ and \textbf{B,} structure factor $S^{zz}_{\rm QPU}(\mathbf{q}_{\rm 1/3})$ for $N=100$ numerically computed using TDVP with maximum bond dimension $D$. Insets showcase the relative error $\varepsilon_{D}$ between data obtained at $D$ and at $D_{\rm max}=700$.}
    \label{fig:mps-scaling}
\end{figure*}

\end{document}